\documentclass{article}

\PassOptionsToPackage{numbers,compress}{natbib}

\usepackage[main, final]{neurips_2026}

\usepackage{amsmath,amssymb}
\usepackage{xcolor}
\usepackage{pifont}
\usepackage{booktabs}
\usepackage[pdftex]{graphicx}

\usepackage{tikz}
\usepackage[most]{tcolorbox}

\usepackage[most]{tcolorbox}
\tcbset{
  colback=white,
  colframe=black,
  boxrule=0.4pt,
  left=6pt, right=6pt,
  top=4pt, bottom=4pt,
  arc=0pt,
  enhanced,
  before skip=8pt, after skip=8pt,
}

\usepackage{xcolor,amssymb}
\usepackage[utf8]{inputenc}
\usepackage[T1]{fontenc}
\usepackage{hyperref}
\usepackage{fontawesome5}
\usepackage{url}
\usepackage{booktabs}
\usepackage{amsfonts}
\usepackage{nicefrac}
\usepackage{microtype}
\usepackage{xcolor}
\usepackage{algorithm}
\usepackage{algpseudocode}

\usepackage{ulem}
\title{A multi-scale information geometry reveals the structure of mutual information in neural populations
}

\author{
  Simone Azeglio \\
  Institut de la Vision, INSERM, CNRS,  \& Laboratoire des Systèmes Perceptifs\\
Sorbonne University \& École Normale Supérieure - PSL \\ 
17 Rue Moreau, Paris 75012 \& 29 Rue d'Ulm Paris 75005\\ \texttt{simone.azeglio@gmail.com}\\
  \And
  Steeve Laquitaine \\
  Institut de la Vision, INSERM, CNRS\\ Sorbonne Université\\ 17 Rue Moreau, Paris 75012 
  \And
  Ulisse Ferrari \\
  Institut de la Vision, INSERM, CNRS\\ Sorbonne Université\\ 17 Rue Moreau, Paris 75012
  \And
  Matthew Chalk \\
  Institut de la Vision, INSERM, CNRS\\ Sorbonne Université\\ 17 Rue Moreau, Paris 75012 \\ \texttt{matthew.chalk@inserm.fr}
}

\begin{document}

\maketitle
\begin{abstract}
Understanding how neural population responses represent sensory information is a central problem in systems neuroscience. One approach is to define a representational geometry on stimulus space in which distances reflect how reliably stimuli can be distinguished from neural activity. However, different constructions of these distances can lead to qualitatively different conclusions about the neural code. Here, we show that a unique Riemannian representational geometry emerges from first principles governing how distances contract as stimulus resolution is lost through coarse-graining. This results in a multi-scale extension of the Fisher information metric, capturing encoding structure from fine stimulus details to coarse global distinctions.
The resulting geometry is exactly related to the mutual information encoded by the population: well-encoded stimulus directions (those contributing more to mutual information) are expanded, whereas poorly encoded directions are contracted.
The metric tensor can be estimated using diffusion models, making the framework practical for large neural populations and high-dimensional stimuli. Applied to visual cortical responses to natural images, the eigenvectors of the metric tensor identify stimulus variations that contribute most to the encoded stimulus information, yielding interpretable features that are robust to modelling choices. Together, these results provide a principled, information-theoretic framework for characterising neural population codes. Code and project page: \href{https://multigeom.github.io/}{multigeom.github.io}~\faGithub
\end{abstract}

\section{Introduction}
How do neural populations represent sensory stimuli? A promising approach is to define a representational geometry over stimulus space, in which distances between stimuli reflect how reliably the population distinguishes them. Such geometries have been used to compare representations across systems and stages of processing~\cite{kriegeskorte2008matching, barbosa2025quantifying}, track changes in representation during learning~\cite{raghu2017svcca, morcos2018insights}, and compare the internal representations of biological and artificial networks~\cite{khaligh2014deep, kornblith2019similarity}. However, many ways exist to define such a geometry from neural responses, and different choices can lead to qualitatively different conclusions about what a population encodes.

Here we show that a unique representational geometry emerges from first principles governing how representational distances transform as stimulus resolution is reduced.
Because neural responses are noisy, we reason that the geometry should depend on the full response distribution rather than treating responses as deterministic. The canonical Riemannian geometry for distributions is constructed from the Fisher information metric, which characterizes how sensitive the response distribution is to infinitesimal stimulus perturbations \cite{amari2016information}. However, Fisher information is purely local: two codes can have identical Fisher information everywhere yet differ markedly in their ability to distinguish distant stimuli~\cite{kriegeskorte2021neural}. We therefore seek a geometry that captures how populations encode stimulus differences at all scales, not only locally. This geometry is uniquely determined by simple requirements on how representational distances change under stimulus coarse-graining, whereby fine stimulus details become progressively harder to discriminate. The resulting geometry provides a multi-scale extension of the Fisher metric, capturing encoding structure from fine discriminations between nearby stimuli to coarse global distinctions.

Our geometry has an exact relationship to the mutual information encoded by the population: distances between stimuli are locally expanded along well-encoded directions, which contribute most to mutual information, and contracted along poorly encoded ones. We show, moreover, that the metric tensor can be estimated for high-dimensional natural stimuli using diffusion models \cite{ho2020denoising, chung2023diffusion}, making the framework practical for neural populations responding to complex natural stimuli.

Our construction builds on identities relating mutual information to the Fisher information, or equivalently the minimum mean squared error, of coarse-grained stimuli across diffusion scales~\cite{guo2005mutual, kong2023information, yu2025mmg, laquitaine2025decomposing}. Previous work has used these identities to decompose mutual information into scalar contributions from individual stimuli or stimulus-response pairs~\cite{laquitaine2025decomposing, kong2023interpretable, dewan2024diffusion}. In contrast, here we define a representational geometry underlying mutual information, whose distances, geodesics and principal directions characterise how information is organised across stimulus space. Moreover, previous decompositions start from mutual information and must choose among many possible ways to decompose it. Instead, we show that our geometry emerges uniquely from simple axioms on how representational distances contract under coarse-graining. The link to mutual information, which is not mentioned in the axioms, thus emerges from the derivation rather than being put in by hand.

We apply our approach to recorded neural populations in visual cortex. The eigenvectors of the metric tensor identify the stimulus directions stretched by the geometry, revealing which stimulus variations contribute most to the mutual information. These directions yield interpretable image features that differ systematically across visual areas. Because the geometry emphasises stimulus directions that contribute to encoded information, and are therefore constrained by the data, these features reflect robust properties of the neural code rather than details of a fitted model. In contrast, we show that Fisher information can be sensitive to stimulus directions that contribute little to encoded information and are therefore poorly constrained. Together, our results establish a principled multi-scale geometry that reveals the structure of encoded information in neural population codes.

\section{From Fisher information to  a multi-scale geometry of neural coding}

Consider a continuous stimulus \(x\sim p(x)\), which elicits a neural population response \(r\) with likelihood \(p(r\mid x)\). Our goal is to define a geometry on stimulus space in which distances reflect how reliably different stimuli can be distinguished from neural activity. This can be formalized by endowing stimulus space with a Riemannian metric, with infinitesimal line element
\begin{equation}
ds^2 = dx^\top G(x)\,dx,
\end{equation}
where \(G(x)\) is some positive semi-definite metric tensor.

A classical result from information geometry, Čencov's theorem \cite{amari2016information}, shows
that under mild conditions on the family of distributions $p(r|x)$ indexed by $x$\footnote{The support of $r$ does not depend on
$x$, and $p(r\mid x)$ is everywhere differentiable in $x$ with finite Fisher.}, the unique metric (up to scale) depending locally on $p(r\mid x)$ and its first derivatives at $x$, while also invariant to transformations of $r$ that preserve its sufficient statistics (i.e.~to reparameterisations that retain all stimulus information), is the Fisher
information matrix $G(x)\propto J(x)$,
\begin{equation}
J(x)
=
\mathbb{E}_{p(r\mid x)}
\left[
\nabla_x \log p(r\mid x)\,
\nabla_x \log p(r\mid x)^\top
\right].
\end{equation} 
The Fisher information matrix thus defines the canonical Riemannian
geometry over the family of response distributions, $p(r\mid x)$. It quantifies how sensitive $p(r\mid x)$ is to infinitesimal
stimulus perturbations through the relation,
$D_\mathrm{KL}(p(r\mid x)\|p(r\mid x+dx))=\tfrac{1}{2}dx^\top J(x)\,dx + O(\|dx\|^3)$.
Geometrically, the Fisher thus deforms stimulus space
so that regions where nearby stimuli elicit more distinguishable responses are stretched further apart.

\begin{figure}
\begin{center}
\includegraphics[width =0.82 \linewidth]{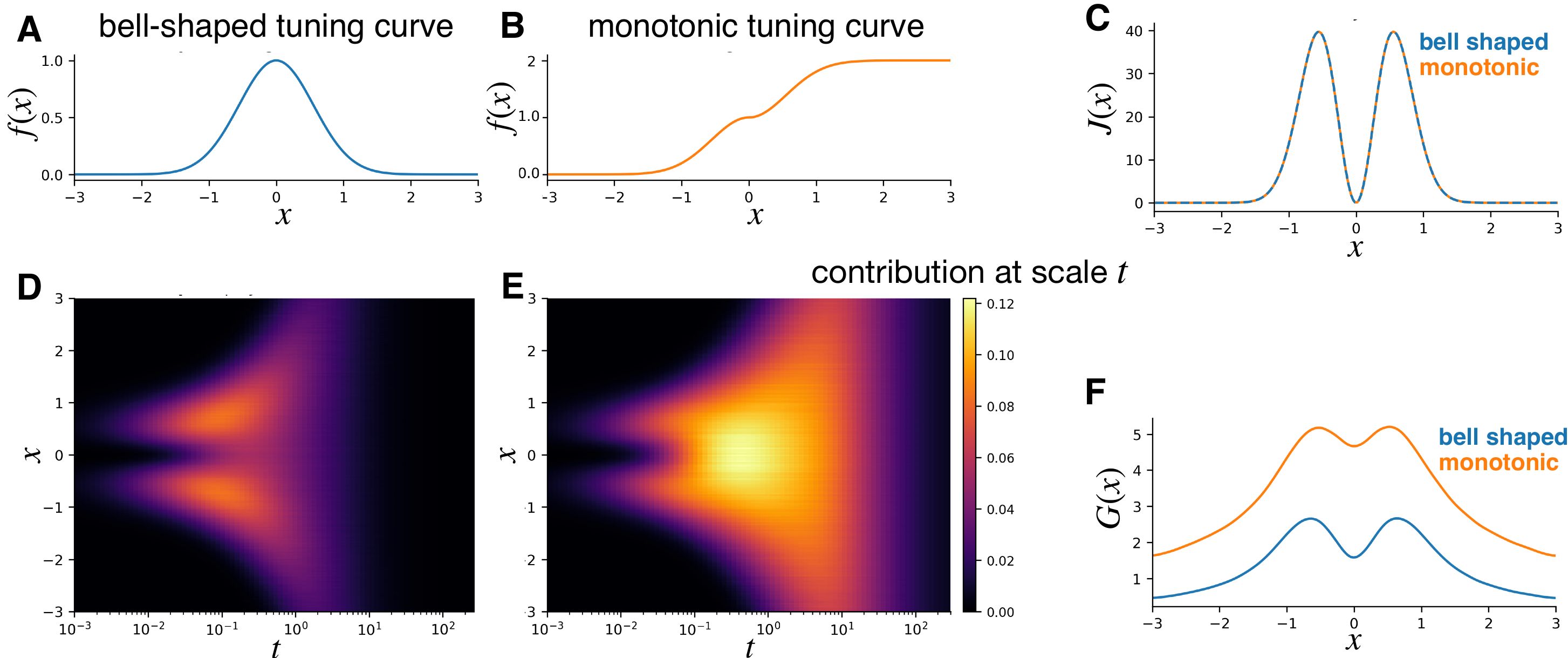}
\caption{\label{fig:1}
\textbf{Fisher and multi-scale Fisher for a one-dimensional stimulus.}
(\textbf{A}--\textbf{B}) Bell-shaped and monotonic tuning curves
($\mathbb{E}[r\mid x]=f(x)$) for a single neuron with
stimulus-independent Gaussian noise, encoding a scalar stimulus $x$
with uniform prior. (\textbf{C}) The Fisher information is identical
for both tuning curves, since they share the same local gradient
magnitude and Gaussian noise variance. (\textbf{D}--\textbf{E})
Contribution at each diffusion scale, $[\phi_t * J_t](x)\,\delta t$,
to the multi-scale Fisher, for the bell-shaped (D) and monotonic (E)
tuning curves. (\textbf{F}) The multi-scale Fisher is larger everywhere
for the monotonic tuning curve, reflecting that it encodes strictly more
mutual information.
}
\end{center}
\end{figure}

However, two neural codes can share identical local structure -- equal
sensitivity to infinitesimal stimulus perturbations -- while differing
arbitrarily in global structure, and thus in how well they distinguish
distant stimuli~\cite{kriegeskorte2021neural}. For example, for a scalar stimulus with Gaussian
observation noise, where $J(x) \propto f'(x)^2$ and $f(x) =
\mathbb{E}[r \mid x]$ is the tuning curve, two tuning curves can share
the same $|f'(x)|$ everywhere, yielding identical Fisher information at
every stimulus, yet encode very different information about the  stimulus globally (Fig.~\ref{fig:1}A--C). This global structure matters because what can be inferred from a response depends on the posterior, $p(x\mid r)\propto p(r\mid x)\,p(x)$, whose normalisation depends on how neurons respond to all stimuli, not only at $x$. Indeed, it is due to its global shape that the monotonic tuning curve (Fig.~\ref{fig:1}B) distinguishes whether $x$ is larger or smaller than zero, while the bell-shaped tuning curve (Fig.~\ref{fig:1}A) does not.

Here we seek a geometry that reflects what the population encodes about the stimulus beyond fine discrimination captured by the Fisher. Requiring the metric itself to depend only on $p(r\mid x)$ and its derivatives at $x$ would force it to be the Fisher information (Čencov's theorem); instead, we impose this locality on how the metric \emph{changes} under coarse-graining. To uniquely define such a geometry, we consider how discriminability is affected by the loss of stimulus resolution induced by coarse-graining with isotropic Gaussian noise:
\begin{equation}
x_t = x + \sqrt{t}\,z,\qquad z\sim\mathcal{N}(0,I). \label{eq:diffn}
\end{equation}
This coarse-graining is the unique choice that is invariant with respect
to translations and rotations of stimulus space, and satisfies the
semigroup property, whereby coarse-graining by $t_1$ and then $t_2$ is
equivalent to coarse-graining once by $t_1+t_2$ \cite{applebaum2009levy}. 

Note that, while we consider an unbounded stimulus space, $x\in\mathbb{R}^{n_x}$, the framework extends to periodic stimuli, such as orientation, by replacing Eq.~\eqref{eq:diffn} with diffusion on the circle (i.e.\ a wrapped Gaussian kernel), which is likewise rotation invariant and satisfies the semigroup property (Appendix~\ref{app:periodic}).

The Gaussian kernel mixes only an $O(\sqrt{t})$ neighbourhood around
$x$, so in the limit $t\to 0$ it reduces the ability of the neural code
to discriminate infinitesimally close stimuli while leaving discrimination
of distant stimuli unaffected. This discrimination of nearby stimuli,
eroded by coarse-graining, depends only on the local shape of
$p(r\mid x)$. We therefore require that (i) distances in our geometry
contract under coarse-graining, since coarse-graining reduces
discriminability, (ii) in the limit $t\to 0$ this contraction depends
locally on $p(r\mid x)$ and its derivatives, and (iii) the contraction depends only on what $r$
encodes about the stimulus, and is thus invariant to
sufficiency-preserving transformations of $r$. Next, we show formally how imposing these requirements on how the
geometry \emph{changes} under coarse-graining, rather than on the metric
itself (as for the Fisher), uniquely determines a metric that integrates
discriminability across stimulus scales.

\section{Deriving the geometry from coarse-graining}\label{sec:axioms}
Under the isotropic Gaussian diffusion of the stimulus described by Eq.~\eqref{eq:diffn}, the joint density 
$p_t(r,x)$ evolves according to the heat equation,
\begin{equation}
\frac{\partial}{\partial t} p_t(r,x) = \frac{1}{2}\Delta_x p_t(r,x),
\end{equation}
with initial condition $p_0(r,x):=p(r\mid x)\,p(x)$. This induces a scale-dependent family of likelihoods
$p_t(r\mid x)=p_t(r,x)/\int p_t(r,x)\,dr$.

Let $G_t(x)$ denote some Riemannian metric on stimulus space constructed from
the diffused joint distribution $p_t(r,x)$. Since coarse-graining
renders $x$ indistinguishable from stimuli within an $O(\sqrt{t})$
neighbourhood, the effective metric at $x$ is the average of $G_t$
over this uncertainty. The rate of change of this
effective metric, evaluated at $t=0$, is
\begin{equation}
Q(x) := \left.\frac{d}{dt}\,\mathbb{E}_{z}\bigl[G_t(x+\sqrt{t}\,z)
\bigr]\right|_{t=0}, \quad z\sim \mathcal{N}(0, I),
 \label{eq:Q_evol}
 \end{equation}
so that $dx^\top Q(x)\,dx$ is the rate of change of the squared line element.
Performing a second-order Taylor expansion of $G_t(x+\sqrt{t}\,z)$, taking the expectation over $z$, and then differentiating
with respect to $t$ (Appendix~\ref{app:derivations}),
\begin{equation}
Q(x)=\left.\left(\frac{\partial}{\partial t}+\frac{1}{2}\Delta_x
\right)G_t(x)\right|_{t=0},\label{eq:Qdef}
\end{equation}
where the Laplacian operator $\Delta_x$ acts component-wise on
$G_t(x)$.

To constrain \(Q(x)\), we impose three requirements:
\begin{enumerate}
\item \textbf{Contraction.} Since coarse-graining reduces what can be inferred from neural responses, local distances should contract, i.e.\ $dx^\top Q(x)\,dx \le 0$ for all $dx$ and $x$. Moreover, this contraction should not depend on the choice of stimulus coordinates.
\item \textbf{Locality.} \(Q(x)\) depends only on \(p(r\mid x)\) and its first derivatives at \(x\).
\item \textbf{Sufficiency.} Contraction under coarse-graining should depend only on what neural responses convey about the stimulus, not on how the response variable is parameterized. Thus, \(Q(x)\) is invariant under any transformation \(r\to r'\) that preserves the sufficient statistics of \(p(r\mid x)\).
\end{enumerate}

Requirement 1, that $Q(x)\preceq 0$ and transforms covariantly under changes of coordinates (Appendix~\ref{app:derivations}), implies that the geometry defined by the metric tensor $G(x)$ is Riemannian. However, Gaussian diffusion cannot be isotropic in every coordinate system, so we define it in one set of base coordinates (image pixels in Section~\ref{sec:v1v4}). Expressed in other coordinates, the geometry is still Riemannian, but the underlying diffusion process is no longer isotropic. The uniqueness of $G(x)$ shown in this section thus holds given a choice of coordinates in which the diffusion process is isotropic Gaussian.

Requirements 2 and 3, together with the positive semi-definiteness of 
$-Q(x)$ (and mild regularity assumptions on $p(r|x)$ stated earlier), are precisely the conditions of Čencov's theorem \cite{amari2016information} (Appendix~\ref{app:derivations}). It follows that
$-Q(x)$ must be proportional to the Fisher information, 
\begin{equation}
-Q(x)=\alpha J(x),
\end{equation}
where \(\alpha>0\) is an arbitrary positive constant. 

This relation holds for any likelihood satisfying the regularity conditions above. Since these are preserved under Gaussian diffusion, we can apply it to the diffused likelihood $p_t(r\mid x)$ at each scale $t$, yielding $-Q_t(x)=\alpha J_t(x)$, where $J_t(x)$ is the Fisher information constructed from $p_t(r\mid x)$. Thus, Eq.~\eqref{eq:Qdef} can be rearranged to give, for all $t\ge 0$,
\begin{equation}
\frac{\partial}{\partial t}G_t(x)= -\frac{1}{2}\Delta_x G_t(x)-\alpha J_t(x).
\label{eq:Qdef_allt}
\end{equation}

This determines the evolution of the metric, but not the metric itself. To uniquely determine \(G(x):=G_0(x)\), we must impose one additional requirement:
\begin{enumerate}
\setcounter{enumi}{3}
\item \textbf{Zero distance baseline.} If \(r\) is independent of \(x\), then \(G(x)=0\) everywhere.
\end{enumerate}
As \(t\to\infty\), the response distribution becomes independent of the stimulus, so that \(\lim_{t\to\infty}p_t(r\mid x)=p(r)\). From the above, this implies
\begin{equation}
\lim_{t\to\infty} G_t(x)=0 \quad \forall x\in \mathcal{X}
\label{eq:boundary}
\end{equation}
Solving Eq.~\eqref{eq:Qdef_allt} backward from the above boundary condition (Appendix~\ref{app:derivations}) then yields a unique solution, up to an overall scaling factor,
\begin{equation}
\boxed{
G(x)=\alpha\int_0^\infty [\phi_t * J_t](x)\,dt,
}
\label{eq:Gdef}
\end{equation}
where \(\phi_t\) is an isotropic Gaussian kernel of variance \(t\), and \(*\) denotes convolution. We henceforth set \(\alpha=1\) without loss of generality, noting that this choice amounts only to an overall scaling of the metric. We call \(G(x)\) the \emph{multi-scale Fisher information matrix}, since it integrates the Fisher information across all coarse-graining scales.

Returning to our previous example, although the Fisher information is identical for both tuning curves (Fig.~\ref{fig:1}A--C), the Fisher information with respect to the \emph{diffused} distribution \(J_t(x)\) differs once \(t>0\) (Fig.~\ref{fig:1}D--E). Consequently, the multi-scale Fisher (Fig.~\ref{fig:1}F) is larger for the monotonic tuning curve, which, as noted earlier, encodes the sign of $x$.

\begin{figure}
\begin{center}
\includegraphics[width =0.92 \linewidth]{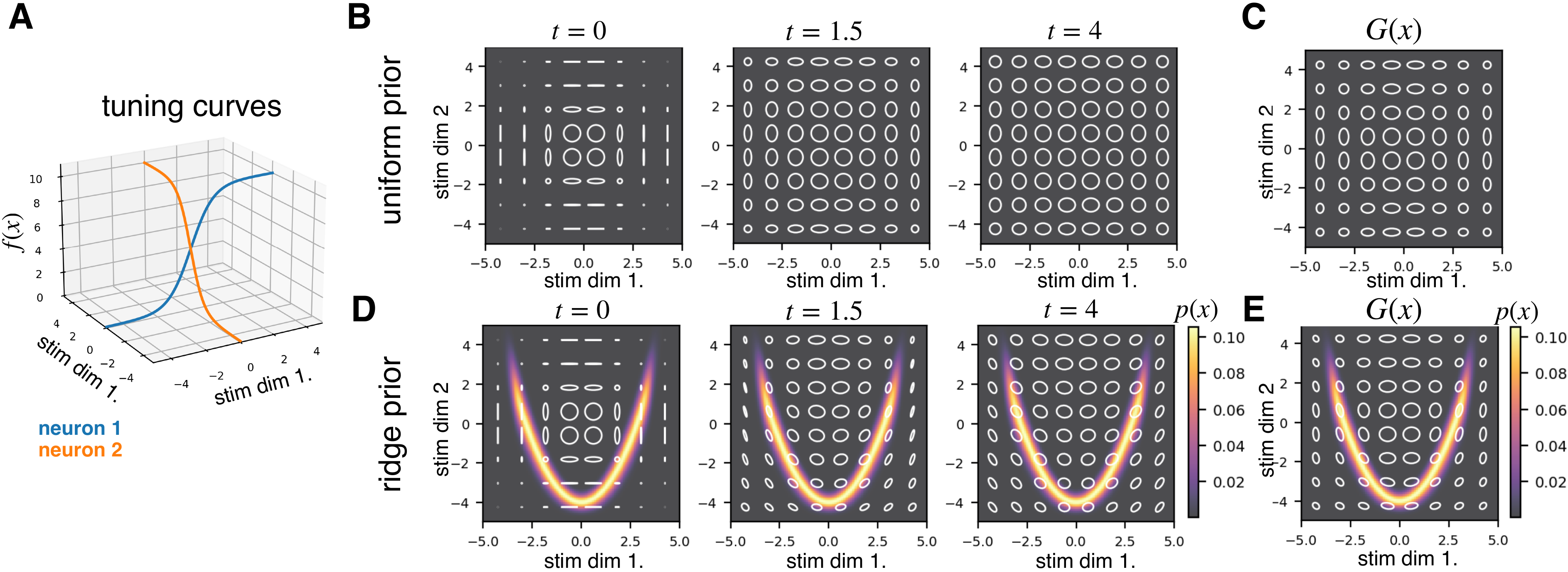}
\caption{\label{fig:2}
\textbf{Multi-scale geometry for a two-dimensional stimulus.}
(\textbf{A}) Tuning curves $f(x)$ for two neurons with
stimulus-independent Gaussian noise, encoding a two-dimensional
stimulus. (\textbf{B}) Contribution at scales $t=0, 1.5$ and $4$ to the
multi-scale Fisher, with uniform prior. Ellipses visualize the local
metric $[\phi_t * J_t](x)\,\delta t$ at equally spaced stimulus values.
(\textbf{C}) Multi-scale Fisher $G(x)$, integrated over all scales.
(\textbf{D}--\textbf{E}) Same as B--C with a non-uniform ridge-shaped
prior (shown in background). For $t>0$, ellipses align with the ridge
of the prior.
}
\end{center}
\end{figure}

To illustrate how the multi-scale geometry depends on both likelihood and prior, we consider a toy example in which two neurons encode a two-dimensional stimulus through sigmoidal tuning curves with Gaussian noise (Fig.~\ref{fig:2}A). We compare a uniform prior (Fig.~\ref{fig:2}B--C) with a prior concentrated along a curved ridge (Fig.~\ref{fig:2}D--E). At $t=0$, the diffused Fisher $J_t(x)$ reduces to the standard Fisher information and is therefore independent of the prior (left panels of Fig.~\ref{fig:2}B,D). For $t>0$, however, the prior enters through the diffused likelihood, $p_t(r\mid x)\propto\int p(r\mid x')\,p(x')\,\phi_t(x-x')\,dx'$ (Appendix~\ref{app:gaussian}), and the diffused Fisher becomes elongated along directions of greatest local stimulus variance (middle and right panels). Integrating over $t$ then yields a multi-scale Fisher that reflects both the likelihood structure and the prior: ellipses are largest near $x=0$, where the tuning curves are steepest, and in the non-uniform case align with the ridge of the prior (Fig.~\ref{fig:2}C,E).

\section{The geometry directly captures mutual information}\label{sec:info}

A central consequence of our construction is that the multi-scale geometry admits an exact mutual-information interpretation. Intuitively, stimulus directions that contribute most to the total mutual information encoded by the population are stretched more strongly by the geometry, pushing nearby stimuli farther apart. Formally, under mild regularity conditions on $p(r, x)$ (Appendix \ref{app:info}),
the metric tensor satisfies \cite{laquitaine2025decomposing}
\begin{align}
\boxed{I(R;X) = \frac{1}{2}\,\mathbb{E}_x \left[\mathrm{Tr}\left(G(x)\right)\right].}\label{eq:infdist1}
\end{align}

This identity has a natural geometric interpretation: mutual information equals the expected rate at which squared distances increase under infinitesimal perturbations of the stimulus,
\begin{align}
I(R;X)=
\frac{1}{2}
\left.
\frac{d}{d\varepsilon}
\mathbb{E}_{x,z}\left[
d^2\left(x; x+\sqrt{\varepsilon}\,z\right)
\right]
\right|_{\varepsilon=0},
\qquad z \sim \mathcal{N}(0, I).\label{eq:infdist2}
\end{align}
Equivalently, for $N$ stimuli sampled from $p(x)$,
\begin{equation}
I(R;X) = \frac{n_x}{2}\lim_{N\to\infty} \frac{1}{N}\sum_{i=1}^N m_i^2, \label{eq:infdist3}
\end{equation}
where $n_x$ is the number of stimulus dimensions, and $m_i = d_i^g / d_i$ is the local magnification at $x_i$: the ratio of the geometric distance $d_i^g$ to the Euclidean distance $d_i$ between $x_i$ and its nearest neighbour (Appendix~\ref{app:info}). Mutual information is thus the mean squared magnification of stimulus space induced by the neural code. Crucially, this identity is not imposed by construction, but emerges as a consequence of our coarse-graining axioms.

This result also clarifies the interpretation of our earlier examples (Figs.~\ref{fig:1}--\ref{fig:2}). For a one-dimensional stimulus with uniform prior, mutual information reduces to the integral of the multi-scale Fisher (Fig.~\ref{fig:1}F). The fact that \(G(x)\) is larger for the monotonic than the bell-shaped tuning curve therefore implies that the monotonic code encodes more mutual information. Likewise, the elongation of \(G(x)\) along directions of greater local stimulus variance (Fig.~\ref{fig:2}E) reflects the stronger contribution of those directions to the total information carried by the population. More  generally, the multi-scale geometry shows how local structure in the likelihood and stimulus prior shapes the encoded information.

\section{Estimating the multi-scale geometry from diffusion models}
\label{sec:estimator}

We can estimate the squared infinitesimal line element $dx^\top G(x)\,dx$ by rewriting Eq.~\eqref{eq:Gdef} as an expectation over $z$ (with $\alpha=1$, following our convention):
\begin{equation}
dx^\top G(x) dx
=
\int_0^\infty
\mathbb{E}_{z\sim \mathcal{N}(0,I)}
\left[
dx^\top J_t(x+\sqrt{t}z)\, dx
\right] dt.
\end{equation}
The integral over $t$ is approximated by the trapezoid rule in $\log t$ (writing $dt = t\,d\log t$) on a discrete grid of scales $\{t_k\}$ and the expectation over $z$ is approximated by Monte Carlo sampling (Appendix~\ref{app:diff-model}).

Direct evaluation of the squared line element, \( dx^\top J_t dx\), is intractable  in high dimensions, since  it requires  integrating over $x$ to obtain the diffused conditional distribution
\(p_t(r \mid x_t) \propto \int \phi_t(x_t-x)\, p(r,x)\, dx\). Instead, we derive an estimator for  \(dx^\top J_t(x_t)dx\) in terms of posterior means that can be learned with a conditional diffusion model.

Expressing $J_t(x_t)$ as the covariance of the score 
$\nabla_{x_t}\log p_t(r\mid x_t)$, and applying Bayes' rule to relate 
the score to $\nabla_{x_t}\log p_t(x_t\mid r)$ (noting that 
$\nabla_{x_t}\log p_t(x_t)$ does not contribute to the covariance),
\begin{equation}
J_t(x_t)
=
\mathrm{Cov}_{p_t(r\mid x_t)}
\!\left[
\nabla_{x_t}\log p_t(x_t\mid r)
\right].
\label{eq:Jt1}
\end{equation}
Under isotropic Gaussian diffusion, Tweedie's identity \cite{efron2011tweedie} gives
\begin{equation}
\nabla_{x_t}\log p_t(x_t\mid r)
=
\frac{1}{t}\bigl(\hat{x}(x_t,r)-x_t\bigr),
\qquad
\hat{x}(x_t,r)=\mathbb{E}[x\mid x_t,r].
\end{equation}
Since \(x_t\) does not depend on \(r\), this yields
\begin{align}
dx^\top J_t(x_t) dx
&=
\frac{1}{t^2}
dx^\top
\mathrm{Cov}_{p_t(r\mid x_t)}[\hat{x}(x_t,r)]
dx \nonumber \\
&=
\frac{1}{2t^2}
\mathbb{E}_{r,r'\sim p_t(r\mid x_t)}
\left[
\bigl((\hat{x}(x_t,r)-\hat{x}(x_t,r'))^\top dx\bigr)^2
\right].
\label{eq:line}
\end{align}
Thus, the diffused Fisher along direction \(dx\) is the expected squared difference between two independently sampled posterior-mean estimates, projected onto \(dx\).

Equation~\eqref{eq:line} expresses \(dx^\top J_t(x_t)dx\) in terms of the posterior mean
\(\hat{x}(x_t,r)=\mathbb{E}[x\mid x_t,r]\), which we estimate using a conditional diffusion model \cite{ho2020denoising, ho2022classifier}. Because the estimator depends only on differences between two samples from the same conditional model, systematic errors cancel; this avoids the mismatch bias of approaches that subtract conditional from separately trained unconditional estimators, which can become potentially unstable as \(t\to 0\) \cite{laquitaine2025decomposing, yu2025mmg, kong2023interpretable, dewan2024diffusion}.

We approximate Eq.~\eqref{eq:line} using Monte Carlo samples \(r, r'\sim p_t(r\mid x_t)\).
Exact sampling would require first drawing \(x\sim p(x\mid x_t)\) and then \(r\sim p(r\mid x)\),
which entails a full backward diffusion pass and is prohibitively expensive. Following
\cite{laquitaine2025decomposing, chung2023diffusion}, we instead use the approximation
\(p_t(r\mid x_t)\approx p(r\mid \hat{x}(x_t))\), where
\(\hat{x}(x_t)=\mathbb{E}[x\mid x_t]\) is obtained in a single forward pass through the
denoising model. This approximation becomes increasingly accurate for small $t$, where the posterior concentrates around its mean. Full implementation details are provided in Appendix~\ref{app:diff-model}.

\section{Uncovering the multi-scale geometry of recorded visual neurons}\label{sec:v1v4}

\begin{figure}
\begin{center}
\includegraphics[width =0.88 \linewidth]{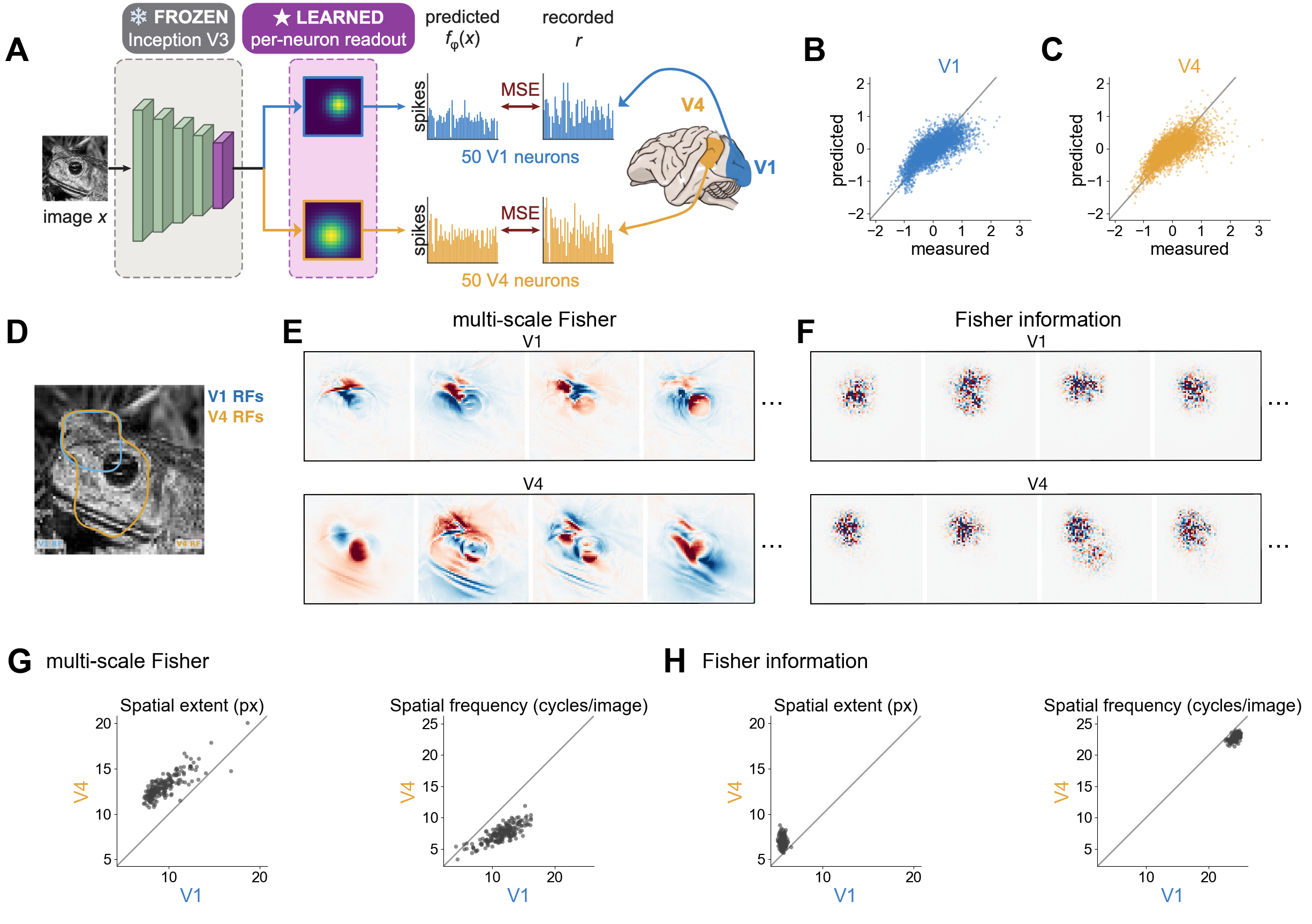}
\caption{\label{fig:3}
\textbf{Multi-scale geometry of neural population recordings in V1 and V4.}
(\textbf{A}) Deep CNN encoding model used to predict V1 and V4 neural
responses to natural images (Appendix~\ref{app:encoder-inception}). (\textbf{B}--\textbf{C})
Measured versus predicted responses across all neurons and images, for
V1 (B) and V4 (C). (\textbf{D}) Example image with V1 and V4 receptive
field regions outlined in blue and yellow, respectively.
(\textbf{E}) Leading four eigenvectors of the multi-scale metric tensor $G(x)$ for this image, for V1 (top) and V4 (bottom). (\textbf{F}) Same as E, for the Fisher information $J(x)$. (\textbf{G}) Top-$10$ eigenvectors of $G(x)$ across the $200$ test images, V1 against V4: spatial extent and spatial frequency (one point per image, averaged over eigenvectors; line, identity). (\textbf{H}) Same as G, for $J(x)$.
}
\end{center}
\end{figure}

We next examined what our multi-scale geometry reveals about how recorded visual populations encode natural images. In practice, computing the geometry requires fitting an encoding model to data, raising a key question: does the resulting geometry reflect intrinsic properties of the neural code, or artifacts of the fitted model?

This issue is particularly acute for high-dimensional stimuli, where encoding models are primarily constrained along directions supported by natural image statistics and remain weakly constrained elsewhere. Because the Fisher information depends directly on local response gradients, it can be sensitive to such poorly constrained directions, leading to model-dependent artifacts.
In contrast, through its connection to mutual information, the multi-scale geometry is expected to emphasize stimulus variations that lie along the natural image manifold and contribute most to the encoded information (Fig.~\ref{fig:2}E). As a result, we hypothesized that it should be more strongly constrained by the data and less sensitive to modeling artifacts.

To test this, we analyzed the dataset of Papale et al.~\cite{papale2025extensive}, consisting of 
large-scale electrophysiological recordings from awake macaque V1 and V4 
during presentation of natural images. We selected the 50 most reliable 
neurons from each area and fit deep encoding models to predict their 
responses from image input (Appendix~\ref{app:encoding_mod}). As is common for encoding models of population recordings~\cite{klindt2017neural, cadena2019deep, lurz2021generalization, willeke2022sensorium}, we modelled trial-to-trial variability as independent Gaussian noise of fixed variance (Eq.~\eqref{eq:obs-model}). The geometries reported here therefore do not account for noise correlations, although the framework itself does not require this assumption (Appendix~\ref{app:corrnoise}). Our  model, based on an 
InceptionV3 architecture \cite{szegedy2016rethinking} (Fig.~\ref{fig:3}A), achieved near state-of-the-art predictive 
performance, with $0.75$ and $0.71$ correlation with responses in V1 and 
V4 respectively (Fig.~\ref{fig:3}B--C).

We first examined the leading eigenvectors of the multi-scale metric tensor $G(x)$ for individual images $x$, which identify local stimulus variations contributing most strongly to encoded information. These eigenvectors differed markedly between cortical areas: in V1 they were spatially localized and edge-like, emphasizing fine image structure, whereas in V4 they were broader and more spatially distributed, capturing more global image structure (Fig.~\ref{fig:3}D--E). Across $200$ held-out TinyImageNet test images, V1 eigenvectors had a smaller spatial extent and higher spatial frequency than V4 eigenvectors (Fig.~\ref{fig:3}G; $p<0.001$, Wilcoxon signed-rank tests; Appendix~\ref{app:eigstats}). In contrast, the leading eigenvectors of the Fisher information resembled high-frequency noise (Fig.~\ref{fig:3}F). Although they also differed significantly between areas (Fig.~\ref{fig:3}H; $p<0.001$; Appendix~\ref{app:eigstats}), these differences were much smaller (mean absolute difference between V1 and V4, for $J(x)$ vs. $G(x)$: $1.4$ vs. $3.7$ pixels in spatial extent, $1.3$ vs. $4.0$ cycles/image in spatial frequency). This is consistent with the Fisher’s sensitivity to response gradients along poorly constrained stimulus directions, which causes its leading structure to be dominated by model-dependent artifacts rather than by intrinsic properties of the neural code.

\begin{figure}
\begin{center}
\includegraphics[width =0.93 \linewidth]{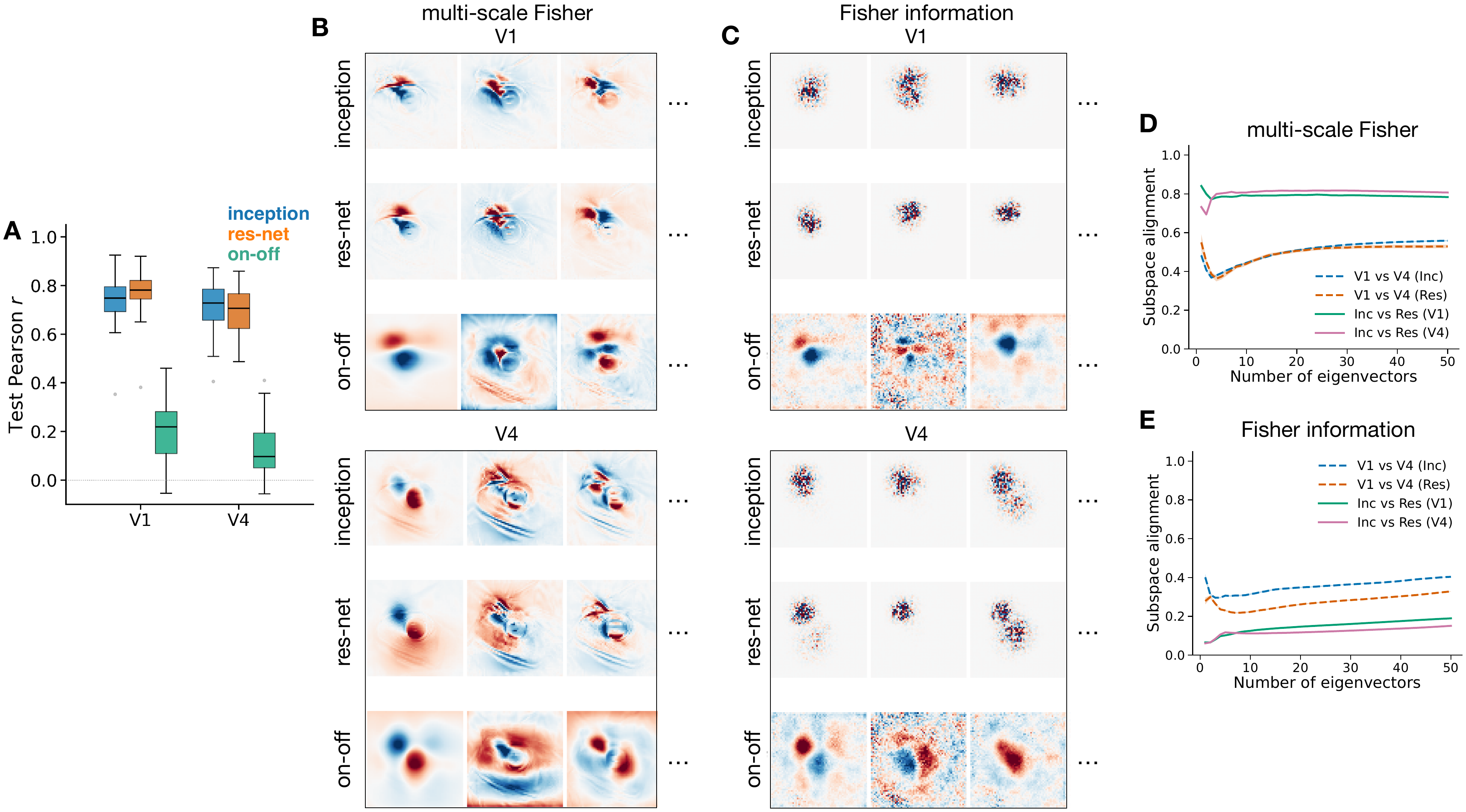}
\caption{\label{fig:4}
\textbf{Robustness across encoding model architectures.}
(\textbf{A}) Predictive performance for Inception, ResNet, and \textsc{On-Off} models, for V1 and V4 (boxplots of test correlation across the $50$ neurons of each area). (\textbf{B}) Leading eigenvectors of the
multi-scale Fisher $G(x)$ for an example image (Fig.~\ref{fig:3}D),
for each model architecture fitted to V1 (top) and V4 (bottom).
(\textbf{C}) Same as B, for the Fisher information $J(x)$.
(\textbf{D}--\textbf{E}) Subspace alignment of the top $k$ eigenvectors
of $G(x)$ (D) and $J(x)$ (E), comparing models fitted to the same area
(solid) versus the same model fitted to different areas (dashed), averaged across images (shading, s.e.m.). }
\end{center}
\end{figure}

We next tested whether this difference was robust across encoding models. 
We fit a second architecture (ResNet-50 \cite{he2016deep}; Appendix~\ref{app:encoding_mod}) with similar predictive 
performance (Fig.~\ref{fig:4}A). The leading eigenvectors of the multi-scale geometry were qualitatively similar across models fit to the same cortical area (Fig.~\ref{fig:4}B). To quantify this, we measured alignment between the subspaces spanned by the top $k$ eigenvectors of each metric tensor, averaged across images. For the multi-scale metric, alignment was higher for two models fit to the same area than for two models of the same 
architecture fit to different areas (Fig.~\ref{fig:4}D; at $k=10$ this held in each of the $200$ images, $p<0.001$, Wilcoxon signed-rank test), indicating that it captured structure intrinsic to the recorded population. For the Fisher information, the ordering was reversed: the two architectures fit to the same area were less aligned with each other than each architecture was with itself across V1 and V4 (Fig.~\ref{fig:4}C \& E; at $k=10$ this held in each of the $200$ images, $p<0.001$, Wilcoxon signed-rank test), implying that its leading eigenvectors are driven more by the choice of encoding model than by the neural data (similar results were obtained using Bures--Wasserstein distance \citep{bhatia2019bures} rather than subspace alignment, Appendix~\ref{app:bures}).

We asked whether the sensitivity of the Fisher geometry to poorly constrained directions could be reduced by using a simpler, less flexible encoding model. We therefore fit neural responses with an ON-OFF model \mbox{\cite{berardino2017eigen}} (Appendix \mbox{\ref{app:encoding_mod}}), whose front-end consists of only $12$ physiologically motivated gain-control parameters, compared with $\sim 173$k frozen convolutional parameters in the InceptionV3 backbone (truncated at \texttt{Conv2d\_4a\_3x3}). The leading Fisher eigenvectors then became more interpretable, revealing edge-like features (Fig.~\mbox{\ref{fig:4}}C). However, predictive performance dropped substantially to $0.20$ and $0.13$ correlation for V1 and V4 respectively (Fig.~\mbox{\ref{fig:4}}A), indicating that the model no longer captured important aspects of the neural code. Thus, meaningful Fisher features could be recovered only by sacrificing model expressiveness, whereas the multi-scale geometry remained interpretable even for high-performing deep CNNs.

\begin{figure}
\begin{center}
\includegraphics[width = 1.0 \linewidth]{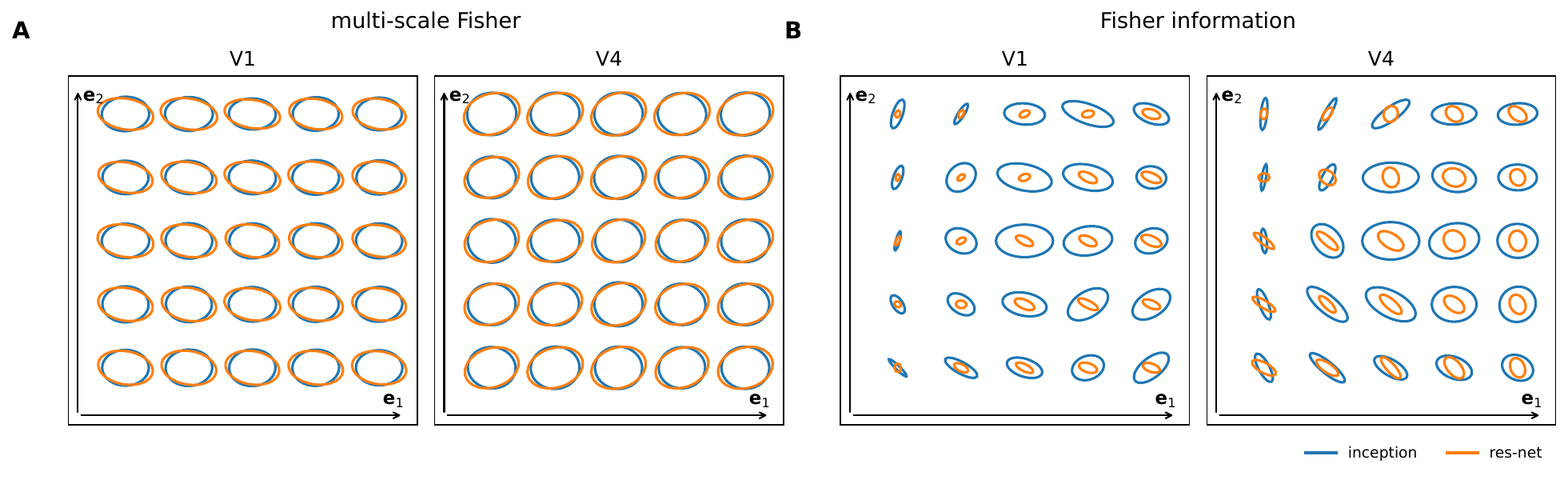}
\caption{\label{fig:5}
\textbf{Local metric field projected onto the leading eigenvectors of
$G(x)$.} (\textbf{A}) Ellipses show the multi-scale
Fisher $G(x)$ computed from Inception (blue) and ResNet (orange)
encoding models, as the reference image is perturbed along the leading
two eigenvectors of $G(x)$ from the Inception model. At each perturbed
image, both metrics are projected onto the same two-dimensional subspace
(leading Inception eigenvectors at the reference image). Results
are shown for V1 (left) and V4 (right). (\textbf{B}) Same as
A, for the Fisher information $J(x)$. 
}
\end{center}
\end{figure}

Finally, we examined how the local geometry evolved when images were 
perturbed along the leading two eigenvectors of $G(x)$ (Fig.~\ref{fig:5}). We 
projected the metric onto the corresponding two-dimensional subspace and 
visualized its local variation as a field of ellipses, with perturbation steps of fixed size, set to $1\%$ of the Euclidean norm of the reference image, along each of the two leading eigenvectors, of $G(x)$ in Fig.~\ref{fig:5}A and of $J(x)$ in Fig.~\ref{fig:5}B, so that both fields cover the same neighbourhood of the reference image. For 
the multi-scale metric, these ellipses changed little across the perturbation space and were nearly identical 
between Inception and ResNet models (Fig.~\ref{fig:5}A). The Fisher ellipses, in contrast, 
varied rapidly and inconsistently across models (Fig.~\ref{fig:5}B), 
confirming that in this setting the Fisher geometry reflects properties of 
the encoding model more than structure in the neural code.

\section{Discussion}\label{sec:discussion}

We derived a Riemannian geometry with an exact relation to mutual information. This connection is rooted in identities relating mutual information to the Fisher information, or equivalently MMSE, across diffusion scales~\mbox{\cite{yu2025mmg, guo2005mutual, kong2023information}}. These have been used to decompose information into contributions from individual stimuli ~\mbox{\cite{laquitaine2025decomposing}} and stimulus-response pairs~\mbox{\cite{kong2023interpretable, dewan2024diffusion}}. Here, instead, we use them to connect \emph{representational geometry} to mutual information. This yields a richer framework than decomposition alone: metric structure, geodesic paths, and distances between stimuli characterize how information is organized across stimulus space, rather than reducing it to scalar contributions from individual stimuli and responses.

Recent work showed that a Riemannian geometry constructed purely from the stimulus distribution $p(x)$ can predict human perceptual judgments~\cite{ohayon2025learning}. In contrast, because our geometry is derived from the full joint distribution $p(r,x)$, it characterizes how well neural populations encode stimuli, not just their statistical structure. It nonetheless yields testable predictions for perception. For any downstream readout $y$ that depends on the stimulus only through $r$, the data-processing inequality for the Fisher information, applied at each scale, implies that distances can only shrink as information passes through successive stages of processing, so that $d_y(x,x')\le d_r(x,x')$. The geometry defined by neural responses should therefore, in principle, constrain perceptual geometry.

A large body of work defines distances between stimuli based on the similarity of elicited neural responses. However, previous works rely on different definitions of representational similarity, under different assumptions: for example, ignoring response variability~\mbox{\cite{raghu2017svcca, kornblith2019similarity}}, relying on mean responses and covariances~\mbox{\cite{schutt2023statistical, williams2021generalized}}, or grounding distances in a deep model's latent space~\mbox{\cite{zhang2018unreasonable, wang2022tuning}}. Here we instead derive a unique, noise-aware geometry from first principles governing how local distances evolve under stimulus coarse-graining. As with mutual information, this avoids arbitrary decoding assumptions and provides a principled description of neural population codes that can be compared across models and datasets.%

Eigenvectors of the Fisher information have been used to identify the stimulus variations best encoded in early visual areas, in shallow networks fitted to neural data~\cite{ding2023information} or in idealized  models (with e.g.~identical, spatially symmetric receptive fields) not fitted to recordings~\cite{berardino2017eigen, feather2024discriminating, zhou2023comparing}. For expressive deep models fitted to V1 and V4, we found the Fisher geometry poorly constrained, with features that vary across architectures, whereas the multi-scale geometry emphasizes directions that are both informative and constrained by the data, and is stable across models fitted to the same data.

A limitation of our approach is that, like the Fisher information, it requires learning a probabilistic encoding model $p(r|x)$ describing how neurons respond to stimuli, and thus sufficient data to estimate this model reliably. Increasingly, however, large-scale datasets of neural responses to natural stimuli are becoming available~\mbox{\cite{papale2025extensive, willeke2022sensorium, turishcheva2024dynamic, schrimpf2018brain, willeke2026omnimouse}}, making such approaches feasible in practice. A second limitation is computational: estimating the geometry is more expensive than the Fisher information, as it requires training a diffusion model and performing multiple evaluations to obtain Monte Carlo estimates of the metric tensor. Nonetheless, diffusion models are widely used, so the approach remains practical despite its increased computational cost.%

Our construction is closely related to renormalization-group (RG) approaches, which describe how systems evolve under coarse-graining~\mbox{\cite{jona1975renormalization}}. Prior work has studied how Fisher information changes along such flows, showing that relevant directions are preserved while irrelevant ones become progressively less distinguishable~\mbox{\cite{raju2018information, maity2015information}}. Our goal here is different: rather than tracking the evolution of a given metric, we derive a unique geometry from first principles and integrate across diffusion scales to obtain a multi-scale metric with an exact link to mutual information. Despite these differences, future work could build on connections to RG theory to better understand how coding structure across scales shapes sensory representations.

\begin{ack}
This work was funded by the Agence Nationale de la Recherche (ANR) ANR-RetNet4EC (to MC), PostGenAI@Paris (ANR-23-IACL-0007 to UF and SA), NatNetNoise (ANR-21-CE37-0024 to UF), IHU FOReSIGHT (ANR-18-IAHU-01 to UF), COCORICO (ANR-26-CE37-4899-01 to UF and MC), and the SCAI through IDEX Sorbonne Université (ANR-11-IDEX-0004, PhD position of SA); Qube Research \& Technologies (QRT), DISSENSATION; and the Conseil Régional d'Île-de-France through DIM C-BRAINS. The authors declare no competing interests.
\end{ack}

\bibliographystyle{unsrtnat}
\bibliography{references_2026,references_cr}

\begin{thebibliography}{52}
\providecommand{\natexlab}[1]{#1}
\providecommand{\url}[1]{\texttt{#1}}
\expandafter\ifx\csname urlstyle\endcsname\relax
  \providecommand{\doi}[1]{doi: #1}\else
  \providecommand{\doi}{doi: \begingroup \urlstyle{rm}\Url}\fi

\bibitem[Kriegeskorte et~al.(2008)Kriegeskorte, Mur, Ruff, Kiani, Bodurka,
  Esteky, Tanaka, and Bandettini]{kriegeskorte2008matching}
Nikolaus Kriegeskorte, Marieke Mur, Douglas~A Ruff, Roozbeh Kiani, Jerzy
  Bodurka, Hossein Esteky, Keiji Tanaka, and Peter~A Bandettini.
\newblock Matching categorical object representations in inferior temporal
  cortex of man and monkey.
\newblock \emph{Neuron}, 60\penalty0 (6):\penalty0 1126--1141, 2008.

\bibitem[Barbosa et~al.(2025)Barbosa, Nejatbakhsh, Duong, Harvey, Brincat,
  Siegel, Miller, and Williams]{barbosa2025quantifying}
Joao Barbosa, Amin Nejatbakhsh, Lyndon Duong, Sarah~E Harvey, Scott~L Brincat,
  Markus Siegel, Earl~K Miller, and Alex~H Williams.
\newblock Quantifying differences in neural population activity with shape
  metrics.
\newblock \emph{bioRxiv}, pages 2025--01, 2025.

\bibitem[Raghu et~al.(2017)Raghu, Gilmer, Yosinski, and
  Sohl-Dickstein]{raghu2017svcca}
Maithra Raghu, Justin Gilmer, Jason Yosinski, and Jascha Sohl-Dickstein.
\newblock {SVCCA}: Singular vector canonical correlation analysis for deep
  learning dynamics and interpretability.
\newblock \emph{Advances in Neural Information Processing Systems}, 30, 2017.

\bibitem[Morcos et~al.(2018)Morcos, Raghu, and Bengio]{morcos2018insights}
Ari~S Morcos, Maithra Raghu, and Samy Bengio.
\newblock Insights on representational similarity in neural networks with
  canonical correlation.
\newblock In \emph{Advances in Neural Information Processing Systems},
  volume~31, pages 5732--5741, 2018.

\bibitem[Khaligh-Razavi and Kriegeskorte(2014)]{khaligh2014deep}
Seyed-Mahdi Khaligh-Razavi and Nikolaus Kriegeskorte.
\newblock Deep supervised, but not unsupervised, models may explain {IT}
  cortical representation.
\newblock \emph{PLoS computational biology}, 10\penalty0 (11):\penalty0
  e1003915, 2014.

\bibitem[Kornblith et~al.(2019)Kornblith, Norouzi, Lee, and
  Hinton]{kornblith2019similarity}
Simon Kornblith, Mohammad Norouzi, Honglak Lee, and Geoffrey Hinton.
\newblock Similarity of neural network representations revisited.
\newblock In \emph{International Conference on Machine Learning}, pages
  3519--3529. PMLR, 2019.

\bibitem[Amari(2016)]{amari2016information}
Shun-ichi Amari.
\newblock \emph{Information geometry and its applications}.
\newblock Springer, 2016.

\bibitem[Kriegeskorte and Wei(2021)]{kriegeskorte2021neural}
Nikolaus Kriegeskorte and Xue-Xin Wei.
\newblock Neural tuning and representational geometry.
\newblock \emph{Nature Reviews Neuroscience}, 22\penalty0 (11):\penalty0
  703--718, 2021.

\bibitem[Ho et~al.(2020)Ho, Jain, and Abbeel]{ho2020denoising}
Jonathan Ho, Ajay Jain, and Pieter Abbeel.
\newblock Denoising diffusion probabilistic models.
\newblock \emph{Advances in Neural Information Processing Systems},
  33:\penalty0 6840--6851, 2020.

\bibitem[Chung et~al.(2023)Chung, Kim, McCann, Klasky, and
  Ye]{chung2023diffusion}
Hyungjin Chung, Jeongsol Kim, Michael~T McCann, Marc~L Klasky, and Jong~Chul
  Ye.
\newblock Diffusion posterior sampling for general noisy inverse problems.
\newblock In \emph{International Conference on Learning Representations}, 2023.

\bibitem[Guo et~al.(2005)Guo, Shamai, and Verd{\'u}]{guo2005mutual}
Dongning Guo, Shlomo Shamai, and Sergio Verd{\'u}.
\newblock Mutual information and minimum mean-square error in gaussian
  channels.
\newblock \emph{IEEE Transactions on Information Theory}, 51\penalty0
  (4):\penalty0 1261--1282, 2005.

\bibitem[Kong et~al.(2023)Kong, Brekelmans, and Steeg]{kong2023information}
Xianghao Kong, Rob Brekelmans, and Greg~Ver Steeg.
\newblock Information-theoretic diffusion.
\newblock In \emph{The Eleventh International Conference on Learning
  Representations}, 2023.

\bibitem[Yu et~al.(2025)Yu, Shi, Kong, Jia, and Steeg]{yu2025mmg}
Longxuan Yu, Xing Shi, Xianghao Kong, Tong Jia, and Greg~Ver Steeg.
\newblock Mmg: Mutual information estimation via the mmse gap in diffusion.
\newblock \emph{arXiv preprint arXiv:2509.20609}, 2025.

\bibitem[Laquitaine et~al.(2025)Laquitaine, Azeglio, Paris, Ferrari, and
  Chalk]{laquitaine2025decomposing}
Steeve Laquitaine, Simone Azeglio, Carlo Paris, Ulisse Ferrari, and Matthew
  Chalk.
\newblock Decomposing stimulus-specific sensory neural information via
  diffusion models.
\newblock In \emph{Advances in Neural Information Processing Systems},
  volume~38, 2025.

\bibitem[Kong et~al.(2024)Kong, Liu, Li, Yogatama, and
  Steeg]{kong2023interpretable}
Xianghao Kong, Ollie Liu, Han Li, Dani Yogatama, and Greg~Ver Steeg.
\newblock Interpretable diffusion via information decomposition.
\newblock In \emph{The Twelfth International Conference on Learning
  Representations}, 2024.

\bibitem[Dewan et~al.(2024)Dewan, Zawar, Saxena, Chang, Luo, and
  Bisk]{dewan2024diffusion}
Shaurya Dewan, Rushikesh Zawar, Prakanshul Saxena, Yingshan Chang, Andrew Luo,
  and Yonatan Bisk.
\newblock Diffusion pid: Interpreting diffusion via partial information
  decomposition.
\newblock \emph{Advances in Neural Information Processing Systems},
  37:\penalty0 2045--2079, 2024.

\bibitem[Applebaum(2009)]{applebaum2009levy}
David Applebaum.
\newblock \emph{L{\'e}vy Processes and Stochastic Calculus}, volume 116 of
  \emph{Cambridge Studies in Advanced Mathematics}.
\newblock Cambridge University Press, Cambridge, 2nd edition, 2009.

\bibitem[Efron(2011)]{efron2011tweedie}
Bradley Efron.
\newblock Tweedie's formula and selection bias.
\newblock \emph{Journal of the American Statistical Association}, 106\penalty0
  (496):\penalty0 1602--1614, 2011.

\bibitem[Ho and Salimans(2021)]{ho2022classifier}
Jonathan Ho and Tim Salimans.
\newblock Classifier-free diffusion guidance.
\newblock In \emph{NeurIPS 2021 Workshop on Deep Generative Models and
  Downstream Applications}, 2021.

\bibitem[Papale et~al.(2025)Papale, Wang, Self, and
  Roelfsema]{papale2025extensive}
Paolo Papale, Feng Wang, Matthew~W Self, and Pieter~R Roelfsema.
\newblock An extensive dataset of spiking activity to reveal the syntax of the
  ventral stream.
\newblock \emph{Neuron}, 113\penalty0 (4):\penalty0 539--553, 2025.

\bibitem[Klindt et~al.(2017)Klindt, Ecker, Euler, and Bethge]{klindt2017neural}
David~A. Klindt, Alexander~S. Ecker, Thomas Euler, and Matthias Bethge.
\newblock Neural system identification for large populations separating
  ``what'' and ``where''.
\newblock In \emph{Advances in Neural Information Processing Systems},
  volume~30, 2017.

\bibitem[Cadena et~al.(2019)Cadena, Denfield, Walker, Gatys, Tolias, Bethge,
  and Ecker]{cadena2019deep}
Santiago~A. Cadena, George~H. Denfield, Edgar~Y. Walker, Leon~A. Gatys,
  Andreas~S. Tolias, Matthias Bethge, and Alexander~S. Ecker.
\newblock Deep convolutional models improve predictions of macaque {V1}
  responses to natural images.
\newblock \emph{PLOS Computational Biology}, 15\penalty0 (4):\penalty0
  e1006897, 2019.
\newblock \doi{10.1371/journal.pcbi.1006897}.

\bibitem[Lurz et~al.(2021)Lurz, Bashiri, Willeke, Jagadish, Wang, Walker,
  Cadena, Muhammad, Cobos, Tolias, Ecker, and Sinz]{lurz2021generalization}
Konstantin-Klemens Lurz, Mohammad Bashiri, Konstantin Willeke, Akshay~K.
  Jagadish, Eric Wang, Edgar~Y. Walker, Santiago~A. Cadena, Taliah Muhammad,
  Erick Cobos, Andreas~S. Tolias, Alexander~S. Ecker, and Fabian~H. Sinz.
\newblock Generalization in data-driven models of primary visual cortex.
\newblock In \emph{International Conference on Learning Representations}, 2021.

\bibitem[Willeke et~al.(2022)Willeke, Fahey, Bashiri, Pede, Burg, Blessing,
  Cadena, Ding, Lurz, Ponder, et~al.]{willeke2022sensorium}
Konstantin~F Willeke, Paul~G Fahey, Mohammad Bashiri, Laura Pede, Max~F Burg,
  Christoph Blessing, Santiago~A Cadena, Zhiwei Ding, Konstantin-Klemens Lurz,
  Kayla Ponder, et~al.
\newblock The sensorium competition on predicting large-scale mouse primary
  visual cortex activity.
\newblock \emph{arXiv preprint arXiv:2206.08666}, 2022.

\bibitem[Szegedy et~al.(2016)Szegedy, Vanhoucke, Ioffe, Shlens, and
  Wojna]{szegedy2016rethinking}
Christian Szegedy, Vincent Vanhoucke, Sergey Ioffe, Jon Shlens, and Zbigniew
  Wojna.
\newblock Rethinking the inception architecture for computer vision.
\newblock In \emph{Proceedings of the IEEE Conference on Computer Vision and
  Pattern Recognition}, pages 2818--2826, 2016.

\bibitem[He et~al.(2016)He, Zhang, Ren, and Sun]{he2016deep}
Kaiming He, Xiangyu Zhang, Shaoqing Ren, and Jian Sun.
\newblock Deep residual learning for image recognition.
\newblock In \emph{Proceedings of the IEEE Conference on Computer Vision and
  Pattern Recognition}, pages 770--778, 2016.

\bibitem[Bhatia et~al.(2019)Bhatia, Jain, and Lim]{bhatia2019bures}
Rajendra Bhatia, Tanvi Jain, and Yongdo Lim.
\newblock On the bures--wasserstein distance between positive definite
  matrices.
\newblock \emph{Expositiones mathematicae}, 37\penalty0 (2):\penalty0 165--191,
  2019.

\bibitem[Berardino et~al.(2017)Berardino, Laparra, Ball{\'e}, and
  Simoncelli]{berardino2017eigen}
Alexander Berardino, Valero Laparra, Johannes Ball{\'e}, and Eero Simoncelli.
\newblock Eigen-distortions of hierarchical representations.
\newblock \emph{Advances in Neural Information Processing Systems}, 30, 2017.

\bibitem[Ohayon et~al.(2026)Ohayon, Fiquet, Guth, Ball{\'e}, and
  Simoncelli]{ohayon2025learning}
Guy Ohayon, Pierre-{\'E}tienne~H. Fiquet, Florentin Guth, Jona Ball{\'e}, and
  Eero~P. Simoncelli.
\newblock Learning a distance measure from the information-estimation geometry
  of data.
\newblock In \emph{The Fourteenth International Conference on Learning
  Representations}, 2026.

\bibitem[Sch{\"u}tt et~al.(2023)Sch{\"u}tt, Kipnis, Diedrichsen, and
  Kriegeskorte]{schutt2023statistical}
Heiko~H Sch{\"u}tt, Alexander~D Kipnis, J{\"o}rn Diedrichsen, and Nikolaus
  Kriegeskorte.
\newblock Statistical inference on representational geometries.
\newblock \emph{Elife}, 12:\penalty0 e82566, 2023.

\bibitem[Williams et~al.(2021)Williams, Kunz, Kornblith, and
  Linderman]{williams2021generalized}
Alex~H Williams, Erin Kunz, Simon Kornblith, and Scott Linderman.
\newblock Generalized shape metrics on neural representations.
\newblock \emph{Advances in Neural Information Processing Systems},
  34:\penalty0 4738--4750, 2021.

\bibitem[Zhang et~al.(2018)Zhang, Isola, Efros, Shechtman, and
  Wang]{zhang2018unreasonable}
Richard Zhang, Phillip Isola, Alexei~A Efros, Eli Shechtman, and Oliver Wang.
\newblock The unreasonable effectiveness of deep features as a perceptual
  metric.
\newblock In \emph{Proceedings of the IEEE Conference on Computer Vision and
  Pattern Recognition}, pages 586--595, 2018.

\bibitem[Wang and Ponce(2022)]{wang2022tuning}
Binxu Wang and Carlos~R Ponce.
\newblock Tuning landscapes of the ventral stream.
\newblock \emph{Cell Reports}, 41\penalty0 (6), 2022.

\bibitem[Ding et~al.(2023)Ding, Lee, Melander, Sivulka, Ganguli, and
  Baccus]{ding2023information}
Xuehao Ding, Dongsoo Lee, Joshua Melander, George Sivulka, Surya Ganguli, and
  Stephen Baccus.
\newblock Information geometry of the retinal representation manifold.
\newblock \emph{Advances in Neural Information Processing Systems},
  36:\penalty0 44310--44322, 2023.

\bibitem[Feather et~al.(2025)Feather, Lipshutz, Harvey, Williams, and
  Simoncelli]{feather2024discriminating}
Jenelle Feather, David Lipshutz, Sarah~E Harvey, Alex~H Williams, and Eero~P
  Simoncelli.
\newblock Discriminating image representations with principal distortions.
\newblock In \emph{The Thirteenth International Conference on Learning
  Representations}, 2025.

\bibitem[Zhou et~al.(2024)Zhou, Chun, Subramanian, and
  Simoncelli]{zhou2023comparing}
Jingyang Zhou, Chanwoo Chun, Ajay Subramanian, and Eero~P Simoncelli.
\newblock Comparing neural models using their perceptual discriminability
  predictions.
\newblock In \emph{Proceedings of UniReps: the First Workshop on Unifying
  Representations in Neural Models}, volume 243, pages 170--181. PMLR, 2024.

\bibitem[Turishcheva et~al.(2024)Turishcheva, Fahey, Vystr{\v{c}}ilov{\'a},
  Hansel, Froebe, Ponder, Qiu, Willeke, Bashiri, Wang,
  et~al.]{turishcheva2024dynamic}
Polina Turishcheva, Paul~G Fahey, Michaela Vystr{\v{c}}ilov{\'a}, Laura Hansel,
  Rachel Froebe, Kayla Ponder, Yongrong Qiu, Konstantin~F Willeke, Mohammad
  Bashiri, Eric Wang, et~al.
\newblock The dynamic sensorium competition for predicting large-scale mouse
  visual cortex activity from videos.
\newblock \emph{ArXiv}, pages arXiv--2305, 2024.

\bibitem[Schrimpf et~al.(2018)Schrimpf, Kubilius, Hong, Majaj, Rajalingham,
  Issa, Kar, Bashivan, Prescott-Roy, Geiger, Schmidt, Yamins, and
  DiCarlo]{schrimpf2018brain}
Martin Schrimpf, Jonas Kubilius, Ha~Hong, Najib~J. Majaj, Rishi Rajalingham,
  Elias~B. Issa, Kohitij Kar, Pouya Bashivan, Jonathan Prescott-Roy, Franziska
  Geiger, Kailyn Schmidt, Daniel L.~K. Yamins, and James~J. DiCarlo.
\newblock Brain-score: Which artificial neural network for object recognition
  is most brain-like?
\newblock \emph{bioRxiv}, page 407007, 2018.

\bibitem[Willeke et~al.(2026)Willeke, Turishcheva, Gilbert, Chakrabarty, Bedel,
  Fahey, Qiu, Weis, Vystr{\v{c}}ilov{\'a}, Muhammad,
  et~al.]{willeke2026omnimouse}
Konstantin~F Willeke, Polina Turishcheva, Alex Gilbert, Goirik Chakrabarty,
  Hasan~A Bedel, Paul~G Fahey, Yongrong Qiu, Marissa~A Weis, Michaela
  Vystr{\v{c}}ilov{\'a}, Taliah Muhammad, et~al.
\newblock Omnimouse: Scaling properties of multi-modal, multi-task brain models
  on 150b neural tokens.
\newblock \emph{arXiv preprint arXiv:2604.18827}, 2026.

\bibitem[Jona-Lasinio(1975)]{jona1975renormalization}
G~Jona-Lasinio.
\newblock The renormalization group: A probabilistic view.
\newblock \emph{Il Nuovo Cimento B (1971-1996)}, 26\penalty0 (1):\penalty0
  99--119, 1975.

\bibitem[Raju et~al.(2018)Raju, Machta, and Sethna]{raju2018information}
Archishman Raju, Benjamin~B Machta, and James~P Sethna.
\newblock Information loss under coarse graining: A geometric approach.
\newblock \emph{Physical Review E}, 98\penalty0 (5):\penalty0 052112, 2018.

\bibitem[Maity et~al.(2015)Maity, Mahapatra, and Sarkar]{maity2015information}
Reevu Maity, Subhash Mahapatra, and Tapobrata Sarkar.
\newblock Information geometry and the renormalization group.
\newblock \emph{Physical Review E}, 92\penalty0 (5):\penalty0 052101, 2015.

\bibitem[{\v{C}}encov(1982)]{cencov1982statistical}
N.~N. {\v{C}}encov.
\newblock \emph{Statistical Decision Rules and Optimal Inference}, volume~53 of
  \emph{Translations of Mathematical Monographs}.
\newblock American Mathematical Society, 1982.

\bibitem[Averbeck et~al.(2006)Averbeck, Latham, and Pouget]{averbeck2006neural}
Bruno~B. Averbeck, Peter~E. Latham, and Alexandre Pouget.
\newblock Neural correlations, population coding and computation.
\newblock \emph{Nature Reviews Neuroscience}, 7\penalty0 (5):\penalty0
  358--366, 2006.
\newblock \doi{10.1038/nrn1888}.

\bibitem[Moreno-Bote et~al.(2014)Moreno-Bote, Beck, Kanitscheider, Pitkow,
  Latham, and Pouget]{moreno2014information}
Rub{\'e}n Moreno-Bote, Jeffrey Beck, Ingmar Kanitscheider, Xaq Pitkow, Peter
  Latham, and Alexandre Pouget.
\newblock Information-limiting correlations.
\newblock \emph{Nature Neuroscience}, 17\penalty0 (10):\penalty0 1410--1417,
  2014.
\newblock \doi{10.1038/nn.3807}.

\bibitem[Le and Yang(2015)]{Le2015TinyIV}
Ya~Le and Xuan~S. Yang.
\newblock Tiny imagenet visual recognition challenge.
\newblock 2015.
\newblock URL \url{https://api.semanticscholar.org/CorpusID:16664790}.

\bibitem[Hebart et~al.(2019)Hebart, Dickter, Kidder, Kwok, Corriveau,
  Van~Wicklin, and Baker]{hebart2019things}
Martin~N Hebart, Adam~H Dickter, Alexis Kidder, Wan~Y Kwok, Anna Corriveau,
  Caitlin Van~Wicklin, and Chris~I Baker.
\newblock Things: A database of 1,854 object concepts and more than 26,000
  naturalistic object images.
\newblock \emph{PloS one}, 14\penalty0 (10):\penalty0 e0223792, 2019.

\bibitem[Yamins and DiCarlo(2016)]{yamins2016using}
Daniel~LK Yamins and James~J DiCarlo.
\newblock Using goal-driven deep learning models to understand sensory cortex.
\newblock \emph{Nature Neuroscience}, 19\penalty0 (3):\penalty0 356--365, 2016.

\bibitem[maintainers and contributors(2016)]{torchvision2016}
TorchVision maintainers and contributors.
\newblock Torchvision: Pytorch's computer vision library.
\newblock \url{https://github.com/pytorch/vision}, 2016.

\bibitem[Duong et~al.(2023)Duong, Bonnen, Broderick, Fiquet, Parthasarathy,
  Yerxa, Zhao, and Simoncelli]{duong2023plenoptic}
Lyndon Duong, Kathryn Bonnen, William Broderick, Pierre-{\'E}tienne Fiquet,
  Nikhil Parthasarathy, Thomas Yerxa, Xinyuan Zhao, and Eero Simoncelli.
\newblock Plenoptic: A platform for synthesizing model-optimized visual
  stimuli.
\newblock \emph{Journal of Vision}, 23\penalty0 (9):\penalty0 5822--5822, 2023.

\bibitem[Golub and Van~Loan(2013)]{golub2013matrix}
Gene~H Golub and Charles~F Van~Loan.
\newblock \emph{Matrix computations}.
\newblock JHU press, 2013.

\bibitem[Harvey et~al.(2024)Harvey, Lipshutz, and
  Williams]{harvey2024representational}
Sarah~E Harvey, David Lipshutz, and Alex~H Williams.
\newblock What representational similarity measures imply about decodable
  information.
\newblock In \emph{Proceedings of UniReps: the Second Edition of the Workshop
  on Unifying Representations in Neural Models}, volume 285, pages 140--151.
  PMLR, 2024.

\end{thebibliography}

\newpage
\appendix

\section{Mathematical details} 

\subsection{Derivations for Section~\ref{sec:axioms}}
\label{app:derivations}

\paragraph{Taylor expansion (Eq.~\eqref{eq:Qdef}).}
Expanding each entry of $G_t$ to second order around $x$,
\begin{equation*}
G_t(x+\sqrt{t}\,z)=G_t(x)+\sqrt{t}\sum_i z_i\,\partial_i G_t(x)+\frac{t}{2}\sum_{i,j}z_iz_j\,\partial_i\partial_j G_t(x)+O(t^{3/2}).
\end{equation*}
Averaging over $z$, using $\mathbb{E}[z_i]=0$ and  $\mathbb{E}[z_i z_j]=\delta_{ij}$, 
\begin{equation}
G_t(x+\sqrt{t}z) = G_t(x)  + \frac{t}{2}  \Delta_x G_t(x) +O(t^{3/2}),
\end{equation}
where $\Delta$ is the Laplace operator, applied to each component of $G_t(x)$. Taking the derivative with respect to $t$, then setting $t=0$ gives,
\begin{equation}
\frac{d}{dt} \mathbb{E}_z[G_t(x+\sqrt{t}z)]\Bigg|_{t=0} = \left. \partial_tG_t(x)\right|_{t=0}   + \left.\frac12 \Delta_x G_t(x)\right|_{t=0} 
\end{equation}
as in Eq.~\eqref{eq:Qdef} in the main text.

\paragraph{Čencov’s theorem.}
Let $\{p(r\mid x)\}$ be a family of distributions whose support does not depend on $x$, differentiable in $x$ with finite Fisher information. The only Riemannian metric on this family that (i) depends only on $p(r\mid x)$ and its first derivatives at $x$, and (ii) is invariant under transformations of $r$ that preserve the sufficient statistics of $p(r\mid x)$ (Markov morphisms), is the Fisher information, up to a positive constant~\cite{cencov1982statistical, amari2016information}.

Requirement 1 states that $dx^\top Q(x)\,dx \le 0$ for all $x$ and $dx$, so $-Q(x)$ is positive semi-definite; $Q(x)$ is symmetric since $G(x)$ is. It also requires the contraction rate to be the same in any coordinates. Let $y=\psi(x)$ be a smooth reparametrisation with Jacobian $\Lambda(x)=\partial\psi/\partial x$, so that $dy=\Lambda\,dx$, and let $\tilde Q(y)$ be the contraction computed in $y$-coordinates for the same coarse-graining process. Then $dy^\top\tilde Q\,dy=dx^\top Q\,dx$ for all $dx$ holds if and only if $Q=\Lambda^\top\tilde Q\,\Lambda$, the transformation law of a covariant rank-2 tensor. Together with the regularity conditions stated in the main text, $-Q(x)$ is therefore a Riemannian metric tensor.

Requirement 2 is condition (i), that $Q(x)$ depends only on $p(r\mid x)$ and its first derivatives at $x$, and Requirement 3 is condition (ii), that $Q(x)$ is invariant under transformations of $r$ that preserve the sufficient statistics of $p(r\mid x)$. All the conditions of Čencov’s theorem are therefore met, and $-Q(x)=\alpha J(x)$ with $\alpha\ge 0$. We exclude the trivial case $\alpha=0$, which gives $G\equiv 0$.

\paragraph{Backward solution (Eq.~\eqref{eq:Gdef}).}
Let $\phi_t*G_t$ denote the convolution of each entry of $G_t$ with the isotropic Gaussian kernel of variance $t$. By the product rule, \[\partial_t(\phi_t*G_t)=(\partial_t\phi_t)*G_t+\phi_t*(\partial_tG_t).\] The heat kernel satisfies $\partial_t\phi_t=\frac12\Delta_x\phi_t$, and differentiation commutes with convolution, $(\Delta_x\phi_t)*G_t=\phi_t*(\Delta_xG_t)$, so, using Eq.~\eqref{eq:Qdef_allt},
\begin{eqnarray*}
\partial_t(\phi_t*G_t)&=&\phi_t*\big(\tfrac12\Delta_xG_t+\partial_tG_t\big)\\&=&-\alpha\,\phi_t*J_t .
\end{eqnarray*}
Integrating from $t=0$ to $\infty$ gives \[ [\phi_t*G_t]_0^\infty=-\alpha\int_0^\infty\phi_t*J_tdt .\] By Eq.~\eqref{eq:boundary} the upper limit vanishes, and $\phi_0*G_0=G_0$, so 
\[ G(x)\equiv G_0(x)=\alpha\int_0^\infty[\phi_t*J_t](x)\,dt, \]
as stated in the text.

\subsection{Relation between geometry and mutual information\label{app:info}}
Recently, \citet{laquitaine2025decomposing} showed that for a memoryless channel $x\rightarrow r$, and assuming that $p(x)$ and $p(x|r)$ have finite first and second moments (for all $r$), the mutual information can be written:
\begin{equation}
I(R;X) = \frac{1}{2}\mathbb{E}_x\left[\mathrm{Tr}\int_{0}^{\infty} (J_t *\phi_t)(x) \, dt\right]
\end{equation}
where $\phi_t$ is an isotropic Gaussian kernel of variance $t$, $*$ denotes convolution, and $J_t$ is the Fisher information computed with respect to the diffused likelihood $p_t(r|x)$. Since the term within the expectation is the trace of the multi-scale Fisher metric $G(x)$, we can write
\begin{equation}
I(R;X) = \frac{1}{2}\mathbb{E}_x\left[\mathrm{Tr}\left(G(x)\right)\right].
\end{equation}
We now show that this is equivalent to Eq.~\eqref{eq:infdist2} in the main text.
For a Riemannian metric, the squared geodesic distance between nearby points satisfies $d^2(x,x+\delta)=\delta^\top G(x)\,\delta+o(\|\delta\|^2)$. With $\delta=\sqrt{\varepsilon}z$, this gives $d^2=\varepsilon\,z^\top G(x)z+o(\varepsilon)$, and the $o(\varepsilon)$ term does not contribute to the derivative at $\varepsilon=0$. Using the local second-order expansion of the squared geodesic distance between $x$ and $x+\sqrt{\varepsilon}z$, where $z\sim \mathcal{N}(0, I)$, $z$ is statistically independent of $x$, and $\varepsilon\to 0$,
\begin{equation}
\frac{1}{2}
\left.
\frac{d}{d\varepsilon}
\mathbb{E}_{x,z}\left[
d^2\!\left(x,\; x+\sqrt{\varepsilon}\,z\right)
\right]
\right|_{\varepsilon=0}
=
\frac{1}{2}
\left.
\frac{d}{d\varepsilon}
\,\varepsilon\, \mathbb{E}_{x,z}\left[
z^\top G(x)\, z
\right]
\right|_{\varepsilon=0} = \frac{1}{2}
\mathbb{E}_{x,z}\left[
z^\top G(x)\, z
\right].
\end{equation}
Applying the identity $z^\top A z = \mathrm{Tr}(A zz^\top)$ and using independence of $z$ and $x$,
\begin{eqnarray}
\frac{1}{2}
\left.
\frac{d}{d\varepsilon}
\mathbb{E}_{x,z}\left[
d^2\!\left(x,\; x+\sqrt{\varepsilon}\,z\right)
\right]
\right|_{\varepsilon=0}
&=&
\frac{1}{2}\,\mathrm{Tr}\!\Big(\mathbb{E}_x\!\left[G(x)\right]\mathbb{E}_z \left[zz^\top\right]\Big) \\
&=&
\frac{1}{2}\mathbb{E}_x\left[\mathrm{Tr}\left(G(x)\right)\right]
\\&=& I(R;X).
\end{eqnarray}

We now prove Eq.~\eqref{eq:infdist3} in the main text.
Consider drawing $N$ iid samples from the stimulus prior, $\{x_i\}_{i=1}^N \sim p(x)$, and let $NN(i)$ denote the Euclidean nearest neighbour of $x_i$. Define the local magnification induced by the metric:
\begin{equation}
m_i
=
\frac{
d^g(x_i, x_{NN(i)})
}{
\|x_i - x_{NN(i)}\|_2
},
\end{equation}
i.e.\ the ratio of the Riemannian distance between two nearby stimuli to their Euclidean separation. Writing the nearest-neighbour displacement as $x_{NN(i)} - x_i = \rho_i u_i$, where $\rho_i > 0$ and $u_i$ is a unit vector, the squared magnification in the large-$N$ limit is
\begin{equation}
\lim_{N\to \infty}m_i^2 = \frac{\rho_i^2\, u_i^\top G(x_i)\,u_i}{\rho_i^2\, u_i^\top u_i} = u_i^\top G(x_i)\,u_i .
\end{equation}
Here $d^g(x_i,x_{NN(i)})^2=\rho_i^2\,u_i^\top G(x_i)\,u_i+o(\rho_i^2)$, and $\rho_i\to 0$ as $N\to\infty$, so the correction vanishes in the limit. Taking expectations over samples and NN directions,
\begin{align}
\lim_{N\rightarrow\infty }\frac{1}{N}\sum_{i=1}^{N}m_i^2
&=
\mathbb{E}_{x,u}[u^\top G(x)u] \\
&=
\mathrm{Tr}
\!\left(
\mathbb{E}_x
\!\left[
G(x)\,\mathbb{E}_{u|x}[uu^\top]
\right]
\right).
\end{align}
As $N\to\infty$, the sample cloud in any vanishing neighbourhood of $x$ converges locally to a homogeneous Poisson point process with intensity $Np(x)$. By the isotropy of the homogeneous Poisson process in the local tangent space, the nearest-neighbour direction is uniformly distributed on $\mathbb{S}^{n_x-1}$, giving
\begin{equation}
\mathbb{E}_{u|x}[uu^\top] = \frac{1}{n_x}I,
\end{equation}
where $n_x$ is the stimulus dimensionality. Substituting back into the expression for the average squared magnification,
\begin{align}
\frac{n_x}{2}
\lim_{N\to\infty}\frac{1}{N}\sum_{i=1}^N m_i^2
&=
\frac{1}{2}
\mathrm{Tr}\!\big(\mathbb{E}_x[G(x)]\big) \\
&= I(R;X),
\end{align}
which is Eq.~\eqref{eq:infdist3} in the main text.

\section{Gaussian stimuli and linear-Gaussian responses}
\label{app:gaussian}

For a Gaussian stimulus $x\sim\mathcal N(0,\Sigma)$ and responses $r=Ax+\eta$, $\eta\sim\mathcal N(0,\sigma^2I)$, all quantities have closed forms. Given the coarse-grained stimulus $x_t=x+\sqrt t\,z$, the stimulus is Gaussian with mean $M_tx_t$, $M_t=\Sigma(\Sigma+tI)^{-1}$, and covariance $C_t=\Sigma-M_t\Sigma$, so $p_t(r\mid x)=\mathcal N(AM_tx,\,AC_tA^\top+\sigma^2I)$ and
\begin{equation*}
J_t=M_t^\top A^\top\big(AC_tA^\top+\sigma^2I\big)^{-1}AM_t,
\qquad G=\int_0^\infty J_t\,dt ,
\end{equation*}
both constant in $x$. The prior enters through $M_t$ and $C_t$: at $t=0$, $M_0=I$, $C_0=0$ and $J_0=A^\top A/\sigma^2=J$, the prior-free Fisher information. For $t>0$, $J_t$ is weighted towards directions of large prior variance. When $\Sigma$ and $J$ share eigenvectors $e_i$ (eigenvalues $s_i$ and $j_i$), $G=\sum_i\log(1+s_ij_i)\,e_ie_i^\top$, and $\frac12\mathrm{Tr}\,G=\frac12\log|I+\Sigma J|=I(R;X)$. 

For weak signals ($J\to0$), $G$ is linear in $J$: in the eigenbasis of $\Sigma$, $G_{ij}=J_{ij}\,s_is_j\log(s_i/s_j)/(s_i-s_j)$, with $G_{ii}=s_iJ_{ii}$. The leading eigenvectors of $G$ therefore lie along directions that are both well encoded and of high prior variance, as in Fig.~\ref{fig:2}.

\section{Robustness to poorly constrained stimulus directions: a toy example}
\label{app:offmanifold}

To illustrate the argument of Section~\ref{sec:v1v4}, we use the linear-Gaussian setting of Appendix~\ref{app:gaussian} with a two-dimensional stimulus whose prior has variance $1$ along a ``manifold'' direction $u_{\rm on}$ and $0.01$ along the orthogonal direction $u_{\rm off}$ (at $30^\circ$), mimicking the low-dimensional structure of natural images. Ten neurons encode $x$ through $A=2\,a_{\rm on}u_{\rm on}^\top+2\beta\,a_{\rm off}u_{\rm off}^\top$, with fixed orthonormal weight vectors $a_{\rm on},a_{\rm off}\in\mathbb R^{10}$ and $\sigma^2=0.5$. The parameter $\beta$ sets how strongly the neurons encode the off-manifold direction. We computed all quantities exactly, evaluating the integral over $t$ by numerical quadrature.

Increasing $\beta$ barely changes the distribution of responses, because the stimulus hardly varies off the manifold, yet it changes the Fisher information completely (Table~\ref{tab:offmanifold}). For $\beta=1$--$2$, the response distribution changes by $0.002$--$0.02$ nats and the off-manifold direction accounts for $1$--$4\%$ of the response variance. Yet the Fisher information becomes as large as or larger off the manifold than on it, and its leading eigenvector turns off the manifold. The multi-scale metric changes by $5$--$16\%$ and its leading eigenvector stays on the manifold. Off-manifold encoding does add information, and $G$ reflects it in proportion to the stimulus variance in that direction.

\begin{table}[h]
\centering
\caption{Off-manifold toy. KL: divergence of the response distribution $p(r)$ from its value at $\beta=0$. Off-var.: fraction of response variance due to the off-manifold direction. Off/on: $u_{\rm off}^\top Mu_{\rm off}/u_{\rm on}^\top Mu_{\rm on}$ for $M=J$ or $G$. Change: relative change of the trace-normalised matrix from $\beta=0$ (Frobenius norm). Angle: angle of the leading eigenvector from $u_{\rm on}$ (at $\beta=1$ the two eigenvalues of $J$ are equal).}
\label{tab:offmanifold}
\footnotesize
\begin{tabular}{rrrrrrrrrr}
\toprule
$\beta$ & KL (nats) & Off-var. & $J$ off/on & $G$ off/on & $J$ change & $G$ change & $J$ angle & $G$ angle & $I$ (nats)\\
\midrule
$0$ & $0$ & $0\%$ & $0$ & $0$ & $0$ & $0$ & $0^\circ$ & $0^\circ$ & $1.099$\\
$0.5$ & $0.0001$ & $0.25\%$ & $0.25$ & $0.009$ & $0.28$ & $0.013$ & $0^\circ$ & $0^\circ$ & $1.109$\\
$1$ & $0.0015$ & $0.99\%$ & $1.00$ & $0.035$ & $0.71$ & $0.048$ & -- & $0^\circ$ & $1.137$\\
$2$ & $0.021$ & $3.9\%$ & $4.00$ & $0.13$ & $1.13$ & $0.16$ & $90^\circ$ & $0^\circ$ & $1.237$\\
$4$ & $0.23$ & $13.8\%$ & $16.0$ & $0.38$ & $1.33$ & $0.39$ & $90^\circ$ & $0^\circ$ & $1.511$\\
$8$ & $1.65$ & $39.0\%$ & $64.0$ & $0.82$ & $1.39$ & $0.64$ & $90^\circ$ & $0^\circ$ & $2.004$\\
\bottomrule
\end{tabular}
\end{table}

\section{Correlated neural noise}
\label{app:corrnoise}

Nothing in the construction of $G(x)$ assumes conditional independence, and the estimator of Section~\ref{sec:estimator} applies to correlated responses if responses are sampled with the full noise covariance. We tested this on a population of $n=6$ neurons encoding a two-dimensional Gaussian stimulus, $x\sim\mathcal N(0,\Sigma_x)$ with unit variances and covariance $0.3$, through overlapping tuning, $\mathbb{E}[r\mid x]=Ax$, with $A$ drawn once with i.i.d.\ $\mathcal N(0,1)$ entries plus $0.6$ times the population-mean tuning vector (rows $(0.49,0.96)$, $(0.47,-1.16)$, $(1.05,0.59)$, $(-0.39,0.72)$, $(0.51,0.43)$, $(0.17,0.69)$). Responses are $r\sim\mathcal N(Ax,\Sigma)$ with baseline per-neuron variance $\sigma^2=0.3$. We introduced correlations in two ways. After encoding, $\Sigma=\sigma^2[I+c\,uu^\top]$, adding ($c=1,4$) or removing ($c=-0.75$) variance along the unit-norm leading eigenvector $u$ of the signal covariance $A\Sigma_xA^\top$. Before encoding, independent noise was added to the stimulus, $r=A(x+\xi_{\rm in})+\xi_{\rm out}$ with $\xi_{\rm in}\sim\mathcal N(0,fI)$, so that $\Sigma=fAA^\top+\sigma^2I$: the correlations then arise because upstream noise is mapped through the weights that carry the signal, the mechanism described by \citet{ding2023information}. The mutual information is $\frac12\log|I+\Sigma_xA^\top\Sigma^{-1}A|$. We computed $\frac12\langle\mathrm{Tr}\,G\rangle$ with the conditional-pair form of Eq.~\eqref{eq:line}, drawing $r=A\tilde x+L\xi$ with $\Sigma=LL^\top$ at stimuli $\tilde x$ sampled from $p(x\mid x_t)$, and using the exact Gaussian posterior means in place of a trained diffusion model. This simulation thus tests the identity and the correlated sampler independently of any trained diffusion network.

\begin{figure}[h]
\begin{center}
\includegraphics[width=1.0\linewidth]{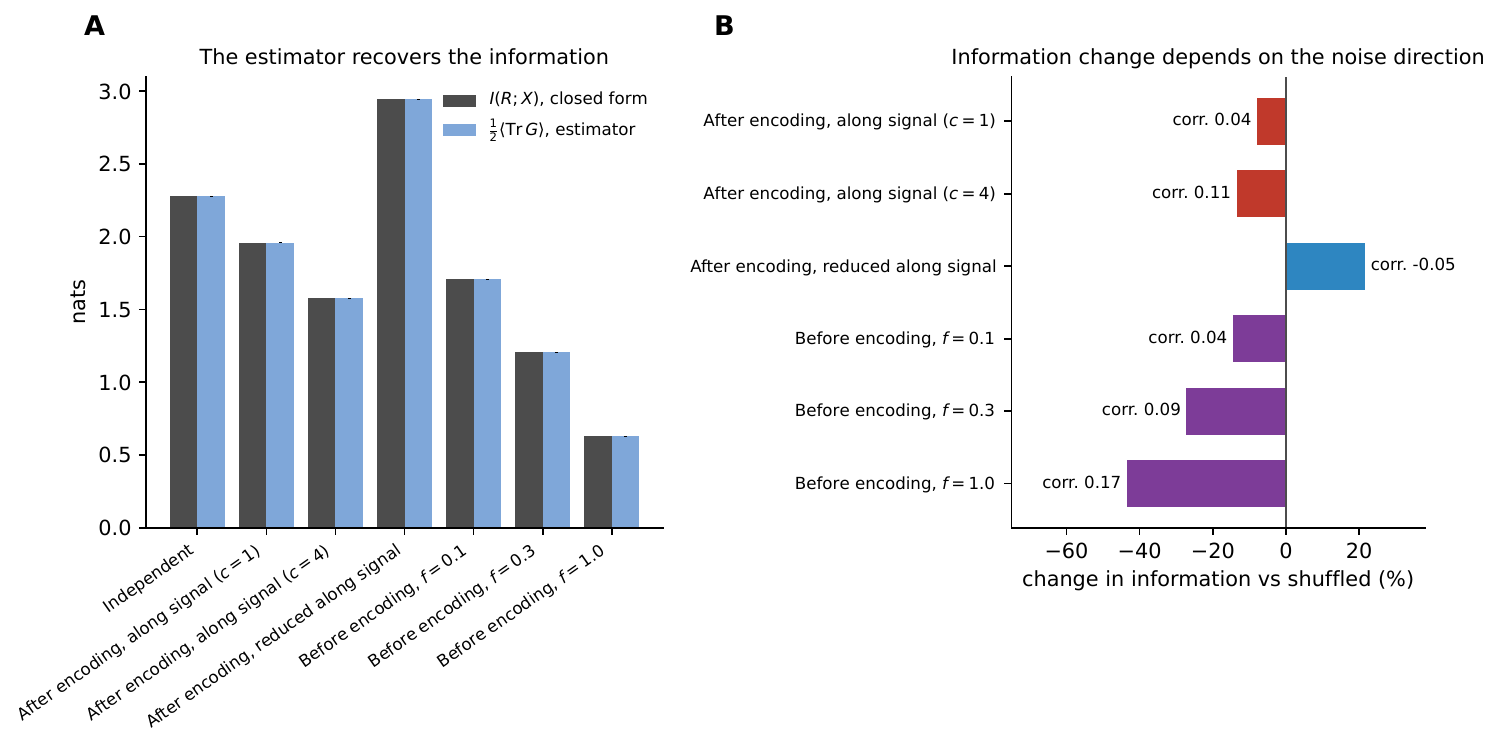}
\caption{\label{fig:corrnoise}
\textbf{The multi-scale geometry under correlated noise.} (\textbf{A}) Mutual information (closed form) and $\frac12\langle\mathrm{Tr}\,G\rangle$ (estimator) for the noise structures of Table~\ref{tab:corrnoise}; the error bars of the estimator (s.e.m.) are smaller than the line width. (\textbf{B}) Change in information relative to the trial-shuffled surrogate for each correlated noise structure, with its mean pairwise noise correlation (corr.). Red, noise added after encoding along the signal direction; blue, noise removed along the signal direction; purple, noise propagated from upstream.}
\end{center}
\end{figure}

The estimate matched the mutual information for every covariance (Table~\ref{tab:corrnoise}, Fig.~\ref{fig:corrnoise}). Relative to a trial-shuffled surrogate with the same single-neuron variances, noise in the signal subspace reduced the information, whether prescribed after encoding ($-7.9\%$ and $-13.5\%$) or propagated from upstream ($-14.5\%$ to $-43.5\%$), and removing noise along the signal direction increased it ($+21.7\%$), as expected from the known effects of noise correlations~\cite{averbeck2006neural, moreno2014information}. Rows~2 and~5 in Table~\ref{tab:corrnoise} have the same mean correlation but differ almost twofold in information loss, so the effect depends on how the correlations are oriented relative to the signal more than on their magnitude.

\begin{table}[h]
\centering
\caption{Correlated noise: mean pairwise noise correlation, mutual information (closed form), $\frac12\langle\mathrm{Tr}\,G\rangle$ (estimator, $\pm$ s.e.m.), their relative difference, and the change in information relative to the trial-shuffled surrogate.}
\label{tab:corrnoise}
\footnotesize
\begin{tabular}{lrrrrr}
\toprule
Noise structure & corr. & $I(R;X)$ & $\frac12\langle\mathrm{Tr}\,G\rangle$ & rel.\ diff. & $\Delta I$ vs shuffled\\
\midrule
Independent (control) & $0.00$ & $2.2774$ & $2.2781\pm0.0011$ & $+0.03\%$ & --\\
After encoding, along signal ($c=1$) & $0.04$ & $1.9596$ & $1.9595\pm0.0009$ & $-0.00\%$ & $-7.9\%$\\
After encoding, along signal ($c=4$) & $0.11$ & $1.5790$ & $1.5796\pm0.0008$ & $+0.04\%$ & $-13.5\%$\\
After encoding, reduced along signal & $-0.05$ & $2.9478$ & $2.9448\pm0.0014$ & $-0.10\%$ & $+21.7\%$\\
Before encoding, $f=0.1$ & $0.04$ & $1.7082$ & $1.7073\pm0.0008$ & $-0.05\%$ & $-14.5\%$\\
Before encoding, $f=0.3$ & $0.09$ & $1.2047$ & $1.2039\pm0.0006$ & $-0.07\%$ & $-27.2\%$\\
Before encoding, $f=1.0$ & $0.17$ & $0.6303$ & $0.6294\pm0.0003$ & $-0.15\%$ & $-43.5\%$\\
\bottomrule
\end{tabular}
\end{table}

\section{Periodic stimuli}
\label{app:periodic}

The requirements of Section~\ref{sec:axioms} concern only the local geometry of stimulus space, so the construction applies on the circle, with coarse-graining by Brownian motion on the circle: $\phi_t$ is the wrapped Gaussian heat kernel and the convolution in Eq.~\eqref{eq:Gdef} is circular. We tested the identity of Section~\ref{sec:info} for populations of Poisson neurons encoding a circular stimulus $x\in[-\pi,\pi)$ with uniform prior. Neuron $i$ has von Mises tuning $f_i(x)=a_i\exp[(\cos(x-\theta_i)-1)/w_i]+0.01$. Rows~1--4 of Table~\ref{tab:periodic} use identical, evenly spaced tuning curves ($a_i=19$, $w_i=0.29$). Rows~5--7 use $10$ neurons with one heterogeneous property (preferred angles displaced by $0.5\sin2\theta_i$; amplitudes from $3.8$ to $34.2$; widths from $0.145$ to $0.58$). Row~8 uses $4$ broad, low-rate neurons ($a_i=5$, $w_i=1.5$), for which the posterior covers much of the circle.

We computed the mutual information, $I(R;X)=H(X)-\mathbb{E}_r[H(X\mid r)]$, by quadrature of each posterior on a $2048$-point grid, averaged over $10^5$ sampled responses. For $\frac12\langle\mathrm{Tr}\,G\rangle=\frac12\int_0^\infty\mathbb{E}_x[J_t(x)]\,dt$, we used $\mathbb{E}_x[J_t]=\mathbb{E}_{r}\big[\mathbb{E}_{p_t(x\mid r)}[(\partial_x\log p_t(r\mid x))^2]\big]$, with $p_t(r\mid x)$ the circular convolution of the likelihood with $\phi_t$ and its derivative computed exactly as Fourier sums, on $39$ log-spaced scales in $[10^{-5},40]$ with the trapezoid rule in $\log t$ (the contribution of $t<10^{-5}$ added from the Fisher information at $t=0$), and independent response samples. The score is computed exactly, without Tweedie's identity, so the test does not involve the Euclidean estimator of Section~\ref{sec:estimator}. With a uniform prior, $p_t(x\mid r)$ is the heat flow of the posterior on the circle, so the identity also holds response by response through de Bruijn's identity. Using independent samples makes the table a test of the numerical integration as well.

\begin{table}[h]
\centering
\caption{Periodic stimulus: mutual information and $\frac12\langle\mathrm{Tr}\,G\rangle$ (nats, $\pm$ s.e.m.) for Poisson populations encoding a circular variable.}
\label{tab:periodic}
\footnotesize
\begin{tabular}{lrrr}
\toprule
Tuning configuration & $I(R;X)$ & $\frac12\langle\mathrm{Tr}\,G\rangle$ & rel.\ diff.\\
\midrule
1 neuron & $0.9240\pm0.0015$ & $0.9227\pm0.0016$ & $-0.13\%$\\
4 neurons & $2.2676\pm0.0008$ & $2.2663\pm0.0008$ & $-0.06\%$\\
10 neurons & $2.8176\pm0.0003$ & $2.8180\pm0.0003$ & $+0.01\%$\\
20 neurons & $3.1680\pm0.0002$ & $3.1676\pm0.0002$ & $-0.01\%$\\
10 neurons, uneven spacing & $2.8017\pm0.0005$ & $2.8026\pm0.0005$ & $+0.03\%$\\
10 neurons, uneven amplitude & $2.7989\pm0.0005$ & $2.7990\pm0.0005$ & $+0.00\%$\\
10 neurons, uneven width & $2.7580\pm0.0005$ & $2.7590\pm0.0005$ & $+0.03\%$\\
4 neurons, broad tuning, low rate & $0.7584\pm0.0010$ & $0.7583\pm0.0010$ & $-0.01\%$\\
\bottomrule
\end{tabular}
\end{table}

\section{Eigenvector statistics across images}
\label{app:eigstats}

\begin{figure}[h]
\begin{center}
\includegraphics[width=1.0\linewidth]{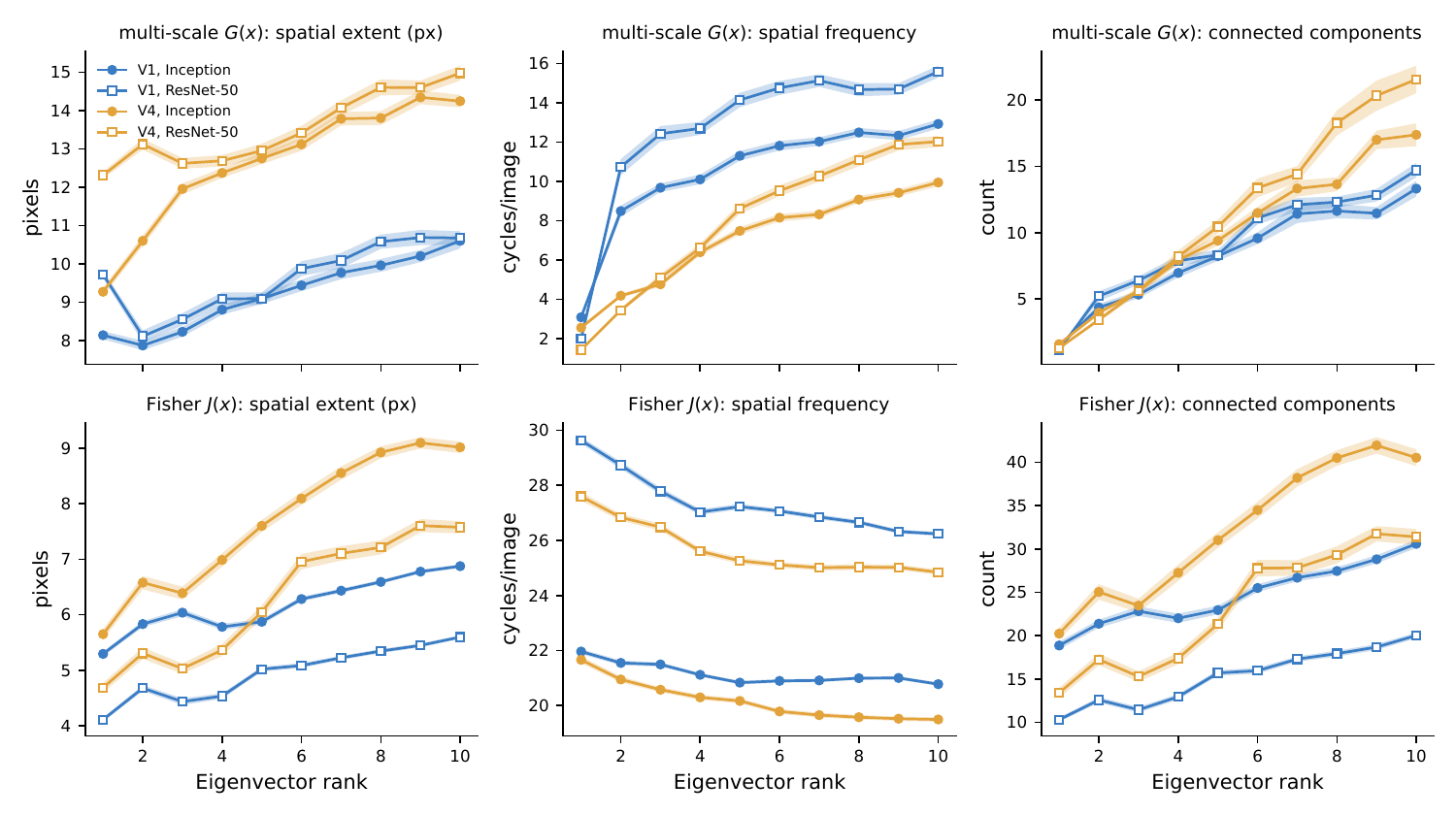}
\caption{\label{fig:eigstats}
\textbf{Eigenvector statistics by rank.} Mean over the $200$ test images (lines) $\pm$ s.e.m. (shading) of the spatial extent, spatial frequency and number of connected components of the $k$-th eigenvector of $G(x)$ (top) and $J(x)$ (bottom), for the Inception (filled circles) and ResNet-50 (open squares) models fitted to V1 (blue) and V4 (yellow).}
\end{center}
\end{figure}

For each of the $200$ test images, each deep encoder (Inception, ResNet-50) and each area, we took the top-$10$ eigenvectors $v$ of $G(x)$ and of $J(x)$ and measured: the spatial extent, the root-mean-square distance of $|v|^2$ from its centroid (pixels); the spatial frequency, the root-mean-square frequency of the power spectrum of $v$ (cycles/image); and the number of connected components of $\{|v|>0.2\max|v|\}$ (the sign of $v$ is ignored). Because the eigenvectors of an image are not independent observations, we averaged each measure over the ranks and the two encoders, and compared V1 and V4 on the same image with a two-sided Wilcoxon signed-rank test across the $200$ images (Fig.~\ref{fig:3}G). For $G(x)$, V1 eigenvectors were smaller (medians $9.0$ vs.\ $12.9$ pixels; $199$ of $200$ images; $p<0.001$), of higher spatial frequency ($11.7$ vs.\ $7.5$ cycles/image; $p<0.001$) and had fewer components ($8.1$ vs.\ $9.5$; $p<0.001$). The extent difference held for each encoder separately  and at every rank. The frequency difference held at all ranks. The difference in components held at thresholds up to $0.5$. Eigenvectors of $J(x)$ also had fewer components in V1, so this measure, like the spatial extent, mainly reflects the smaller receptive fields in V1. For $J(x)$, V1 eigenvectors were also smaller in terms of spatial extent ($5.6$ vs.\ $7.0$ pixels; $p<0.001$), consistent with smaller receptive fields, but the difference in spatial frequency was small ($24.3$ vs.\ $22.9$ cycles/image; $p<0.001$; a ratio of $1.06$, compared with $1.57$ for $G(x)$). Shown separately for each architecture (Fig.~\ref{fig:eigstats}), the difference between areas in $G(x)$ was larger than the difference between architectures (spatial frequency in V1 and V4: $10.6$ and $7.0$ cycles/image for Inception, $12.9$ and $8.0$ for ResNet-50). For $J(x)$, the architectures differed more than the areas ($21.3$ and $20.2$ cycles/image for Inception, $27.3$ and $25.7$ for ResNet-50).

\section{Data efficiency and Monte Carlo convergence}
\label{app:mc}

The only stage limited by recorded data is the encoding model $p(r\mid x)$, which $G(x)$ requires just as the Fisher information does. Our encoders reach $0.81$ and $0.79$ of the noise ceiling in V1 and V4 (Table~\ref{tab:encoder_perf}). Given the encoder, the diffusion model is trained on synthetic pairs from an effectively unlimited image set, and the remaining error is Monte Carlo error. The metric is defined on stimulus space, so its dimension does not grow with the number of neurons. We did not measure how the estimated geometry changes with the amount of recorded data or the number of recorded neurons.

\begin{figure}[h]
\begin{center}
\includegraphics[width=0.85\linewidth]{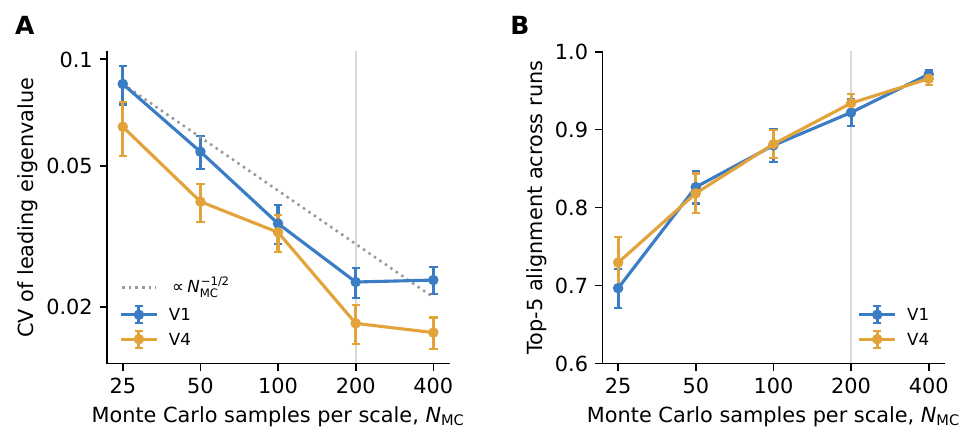}
\caption{\label{fig:mc}
\textbf{Monte Carlo convergence of the estimator} ($10$ test images, $5$ runs each, Inception models; mean $\pm$ s.e.m.\ over images). (\textbf{A}) Coefficient of variation of the leading eigenvalue of $G(x)$ across runs; dotted line, $N_{\rm MC}^{-1/2}$. (\textbf{B}) Mean pairwise alignment of the top-$5$ eigenvectors across runs (Eq.~\eqref{eq:subspace-alignment}). Grey vertical line, $N_{\rm MC}=200$, the value used throughout.}
\end{center}
\end{figure}

To quantify the Monte Carlo error, we ran Algorithm~\ref{alg:cc} five times with independent random seeds for each of $10$ test images, for the Inception models of V1 and V4, at $N_{\rm MC}=25$, $50$, $100$, $200$ and $400$ samples per scale (Fig.~\ref{fig:mc}). The coefficient of variation of the leading eigenvalue across runs fell with $N_{\rm MC}$ up to $N_{\rm MC}=200$, the value used throughout (log-log slope $-0.62$ in V1 and $-0.58$ in V4 between $25$ and $200$ samples, close to the $N_{\rm MC}^{-1/2}$ expected for independent samples), and did not decrease further at $400$ (Fig.~\ref{fig:mc}A). The top-$5$ eigenvectors were reproducible across runs (mean squared cosine of the principal angles $0.92$ in V1 and $0.93$ in V4 at $N_{\rm MC}=200$) and became more so at $400$ ($0.97$; Fig.~\ref{fig:mc}B).

\section{Estimating the geometry using diffusion models\label{app:diff-model}}

This appendix details the diffusion model used to approximate the posterior
mean $\hat{x}(x_t,r)=\mathbb{E}[x\mid x_t,r]$ that appears in
Eq.~\eqref{eq:line}, and the Monte Carlo estimator we employ to obtain the
leading eigenvectors of the multi-scale metric $G(x)$.

\subsection{Diffusion model architecture and training}
\label{app:diff-model-arch}

The goal of this subsection is to specify a conditional score network
$\epsilon_\theta(x_t,t,r)$ from which we can recover the conditional
posterior mean $\hat x(x_t,r)=\mathbb{E}[x\mid x_t,r]$ that enters
Eq.~\eqref{eq:line}.

\paragraph{Forward diffusion.}
The main-text coarse-graining of Eq.~\eqref{eq:diffn} is the
\emph{variance-exploding} process
\begin{equation}
x_\tau \;=\; x \;+\; \sqrt{\tau}\,z,
\qquad z\sim\mathcal{N}(0,I), \quad \tau\geq 0,
\label{eq:forward-ve}
\end{equation}
where $\tau$ is the variance of the added noise; we use $\tau$ here
to distinguish the continuous main-text scale from the discrete DDPM
step $t$ introduced below. To implement Eq.~\eqref{eq:forward-ve}
with standard DDPM tooling, we adopt the \emph{variance-preserving}
parametrisation of \citet{ho2020denoising}: for a clean image
$x\in[-1,1]^{64\times 64}$ (single channel), the noisy variable at
DDPM step $t\in\{1,\ldots,T\}$ is
\begin{equation}
x_t \;=\; \sqrt{\bar\alpha_t}\,x \;+\; \sqrt{1-\bar\alpha_t}\,z,
\qquad z\sim\mathcal{N}(0,I),
\label{eq:forward-diff}
\end{equation}
with $T=1000$ and $\bar\alpha_t$ the cumulative product of the linear
$\beta$-schedule of \citet{ho2020denoising}. The two processes are
equivalent up to a deterministic rescaling: defining
$\tilde x_t := x_t/\sqrt{\bar\alpha_t}$ and substituting
Eq.~\eqref{eq:forward-diff} gives
\begin{equation}
\tilde x_t \;=\; x \;+\; \sqrt{\tfrac{1-\bar\alpha_t}{\bar\alpha_t}}\,z,
\end{equation}
which matches Eq.~\eqref{eq:forward-ve} under the identification
\begin{equation}
\tau \;=\; \frac{1-\bar\alpha_t}{\bar\alpha_t},
\label{eq:time-map}
\end{equation}
so the main-text scale $\tau$ equals the inverse signal-to-noise
ratio at DDPM step $t$, sweeping $(0,\infty)$ monotonically as $t$
runs from $1$ to $T$. Since $x_t$ and $\tilde x_t$ differ only by an
invertible linear transformation, conditioning on either yields the
same posterior, so
$\hat x(x_t,r)=\mathbb{E}[x\mid x_t,r]=\mathbb{E}[x\mid \tilde x_t,r]$
is the same in both parametrisations, and the estimator in
Eq.~\eqref{eq:line} is unchanged. The Jacobian of the
reparametrisation $\tau\mapsto t$ is absorbed by integrating along
$\log\tau$ rather than $\tau$ directly, which is precisely the
log-SNR grid used in Appendix~\ref{app:diff-model-est}.

\paragraph{Score network.}
A conditional score network $\epsilon_\theta(x_t,t,r)$ is trained to
predict the noise realisation $z$ from the noisy image, the
diffusion step and the response vector $r$, using the standard DDPM
mean-squared-error objective
$\mathbb{E}\bigl[\|\epsilon_\theta(x_t,t,r)-z\|^2\bigr]$
\citep{ho2020denoising}. At the minimum of this objective,
$\epsilon_\theta(x_t,t,r)=\mathbb{E}[z\mid x_t,t,r]$. Taking
$\mathbb{E}[\,\cdot\,\mid x_t,r]$ on both sides of
Eq.~\eqref{eq:forward-diff} and using linearity of conditional
expectation then yields the conditional posterior mean as a direct
function of the learned noise prediction,
\begin{equation}
\hat x(x_t,r)
\;=\; \frac{1}{\sqrt{\bar\alpha_t}}\Bigl(x_t \;-\; \sqrt{1-\bar\alpha_t}\,\epsilon_\theta(x_t,t,r)\Bigr)
\;=\; \mathbb{E}[x\mid x_t,r].
\label{eq:xhat-eps}
\end{equation}
The same relation with $r$ replaced by the null token $r=\mathbf 0$
recovers the \emph{unconditional} posterior mean
$\hat x(x_t)=\mathbb{E}[x\mid x_t]$ from the null-token prediction
$\epsilon_\theta(x_t,t,\mathbf 0)$; we exploit this fact in the next
subsection.

\paragraph{Observation model for responses.}
Throughout the paper we treat the recorded response vector $r$ as a
noisy observation of a deterministic image-to-response map $f_\phi(x)$
(the encoder, Appendix~\ref{app:encoding_mod}),
\begin{equation}
p(r\mid x) \;=\; \mathcal{N}\!\bigl(f_\phi(x),\,\sigma^2 I\bigr),
\qquad \sigma=0.22.
\label{eq:obs-model}
\end{equation}
The Gaussian form matches the statistics of the responses we actually
use: the \texttt{normMUA} of \citet{papale2025extensive} is
trial-averaged and per-channel $z$-scored
(Appendix~\ref{app:data}), producing an approximately continuous and
symmetrically distributed signal rather than raw spike counts. This
differs from the setting of \citet{laquitaine2025decomposing}, whose
conditioning vector is the (integer) spike count of a 49-neuron
LNP-style model and is therefore drawn, at every training step, from
a Poisson observation model
$r\sim\mathrm{Poisson}(f_\phi(x))$. Both choices share the same
structure---fix the encoder $f_\phi$ and match the training-time
distribution of $r$ to the observation model consistent with the
data---but differ in which observation model is appropriate for the
signal at hand. We calibrate $\sigma$ to the residual trial-to-trial
variability of the recordings: each held-out stimulus in
\citet{papale2025extensive} is presented $30$ times, and for each
electrode we compute the standard deviation of the $z$-scored
per-trial response around the trial-averaged mean. The median of
this quantity across electrodes is $0.22$ in V1 and $0.25$ in V4; we use $\sigma=0.22$ for both areas, so that a single observation-noise scale is shared across all encoders and areas.

We stress that this $\sigma$ is an actual property of the observation
model in Eq.~\eqref{eq:obs-model}, not merely a regularisation
trick: it appears unchanged in the training recipe below and in the
response sampler of Appendix~\ref{app:diff-model-est}.

\paragraph{Architecture.}
The score network $\epsilon_\theta$ is implemented as a 2D UNet with
cross-attention conditioning, using the standard
\texttt{UNet2DConditionModel} configuration from the Hugging Face
\texttt{diffusers} library. It operates on $64\times 64$ grayscale
images (one input channel, one output channel) with block-output
widths $(64, 128, 256, 512)$, two residual layers per block, and cross-attention layers in the $16\times 16$ down and up blocks and in the $8\times 8$ middle block, following the implementation of
\citet{laquitaine2025decomposing}. Neural responses enter through a
three-layer MLP with SiLU activations that maps the $50$-dimensional
response vector $r$ to a single $256$-dimensional token, which is fed
to the cross-attention layers as the \texttt{encoder\_hidden\_states}
argument. Together with the conditioning embedder, the model has
approximately $68$ million parameters.

\paragraph{Training-time noise injection and classifier-free guidance.}
Two features of the training recipe are essential for the estimator
of Eq.~\eqref{eq:line} to be numerically stable. First, we inject the
observation noise of Eq.~\eqref{eq:obs-model} into the conditioning
vector at every training step: during each forward pass the network
sees $\tilde r = r + \sigma\xi$, with $\xi\sim\mathcal{N}(0,I)$ and
the same $\sigma=0.22$ as in Eq.~\eqref{eq:obs-model}; note that this
observation noise $\xi$ is an independent random variable from the
diffusion-process noise $z$ in Eq.~\eqref{eq:forward-diff}.

Second, we apply classifier-free guidance (CFG) dropout at rate
$p_{\text{drop}}=0.1$ \citep{ho2022classifier}: with that
probability, the entire conditioning vector is replaced by zero
during a training step. A single network therefore learns both the
conditional noise estimate $\epsilon_\theta(x_t,t,r)$ and, when
queried with $r=\mathbf 0$, an unconditional estimate
$\epsilon_\theta(x_t,t,\mathbf 0)\approx\epsilon^{u}_\theta(x_t,t)$.
Without CFG training the model has no meaningful response at the
null token, and the unconditional posterior mean
$\hat x(x_t)=\mathbb{E}[x\mid x_t]$ used in the denoising step of
the next subsection cannot be recovered from a single forward pass.

\paragraph{Optimisation and schedule.}
Training uses the DDPM MSE objective \citep{ho2020denoising} with
$T=1000$ steps on the linear $\beta$-schedule. We optimise with AdamW
at learning rate $10^{-4}$, weight decay $10^{-4}$, and a cosine
schedule with $500$ warm-up steps, for $200$ epochs on $100{,}000$
TinyImageNet~\citep{Le2015TinyIV} training images resized to $64\times 64$ and converted to
single-channel gray. TinyImageNet has no associated neural recordings; the conditioning vector $r$ for each training image is therefore generated synthetically by evaluating the deterministic encoder $f_\phi$ (Appendix~\ref{app:encoding_mod}) on the image, yielding a $50$-dimensional response vector that is then perturbed by the observation noise $\sigma\xi$, $\xi\sim\mathcal{N}(0,I)$, described above. The encoder $f_\phi$ is itself fitted on the macaque V1/V4 \texttt{normMUA} of \citet{papale2025extensive}, so the conditioning distribution seen by the diffusion model approximates that of the recorded population. Batches of $128$ images are processed in
\texttt{bfloat16} precision; we clip gradients
at global norm $1.0$. We use the checkpoints after $200$ epochs throughout. One diffusion model is trained per (cortical
area, encoder) pair in Figs.~\ref{fig:3}--\ref{fig:4}.

\paragraph{Compute resources.}
All training and inference ran on an internal cluster of NVIDIA~A100-SXM4-40GB GPUs. Each conditional diffusion model was trained on a single A100, at $\approx\!4.1$~min per epoch, so a $200$-epoch run takes $\approx\!13.5$ hours. The six models (Inception, ResNet, ON-OFF, each fitted to V1 and V4) cost $\approx\!81$ A100-GPU-hours. The encoder readouts on top of frozen backbones are trained on pre-extracted features and are negligible in comparison ($<1$~A100-GPU-hour in total). Estimating $G(x)$ with Algorithm~\ref{alg:cc}, together with the Fisher baseline, took $33$--$85$~s per image on a single A100 (median $\approx\!40$~s; nodes were shared with other jobs), so the $200$ test images $\times$ $6$ configurations add $\approx\!15$ A100-GPU-hours, for a total of $\approx\!100$ A100-GPU-hours for the reported experiments. The Monte Carlo convergence analysis (Appendix~\ref{app:mc}) used a further $\approx\!8$ A100-GPU-hours, and the simulations of Appendices~\ref{app:gaussian}--\ref{app:periodic} ran on CPUs in minutes. Earlier and exploratory runs outside the reported sweep, including diffusion models trained for $500$ epochs, used more than four times this compute.

\subsection{Estimating metric tensor eigenvectors with the diffusion model}
\label{app:diff-model-est}

The goal of this subsection is to describe the Monte Carlo estimator
used to obtain the leading eigenvectors of the multi-scale metric
$G(x)$ from the trained score network $\epsilon_\theta$ of
Appendix~\ref{app:diff-model-arch}.

\paragraph{Numerical integration over diffusion scales.}
Given a clean image $x$, we estimate $G(x)$ by Monte Carlo integration
of Eq.~\eqref{eq:line} over diffusion scales $t$ and realisations of
the forward-process noise $z\sim\mathcal{N}(0,I)$ in
Eq.~\eqref{eq:forward-diff}. We discretise $t$ on a coarse grid
$\{t_j\}_{j=1}^{J}$ with $t_1=40$, $t_J=960$ and step $40$
($J=24$ timesteps), and for each $t_j$ draw $N_{\text{MC}}=200$
independent noise samples in mini-batches of $25$. The line element
of Eq.~\eqref{eq:line} is smooth in the log signal-to-noise variable
$\log\tau$ (with $\tau=(1-\bar\alpha_t)/\bar\alpha_t$ from
Eq.~\eqref{eq:time-map}) rather than in $t$ itself, so we integrate
with respect to $\log\tau$ using the trapezoid weights
$w_j=\tfrac12(\log\tau_{j+1}-\log\tau_{j-1})$. This is a change
of integration variable: the
timesteps $t_j$ are spaced uniformly on the DDPM index axis, and
$w_j$ is the corresponding discrete integration step $\Delta(\log\tau_j)$ in the log-SNR variable. The Jacobian of the reparametrisation, $\mathrm{d}\tau = \tau\,\mathrm{d}(\log\tau)$, has already been absorbed into the integrand: combining the $1/\tau^2$ factor that Tweedie's identity contributes to $J_t(\tilde x_t)$ in Eq.~\eqref{eq:line} with the relation $\hat x(\tilde x_t, r) - \tilde x_t = -\sqrt{\tau}\,\epsilon_\theta(x_t,t,r)$ leaves a $1/\tau$ factor, which is exactly cancelled by the $\tau$ from $\mathrm{d}\tau = \tau\,\mathrm{d}(\log\tau)$, so the $\log\tau$-integrand reduces to squared noise-prediction differences weighted by $w_j$ (Algorithm~\ref{alg:cc}). The resulting matrix
$D\in\mathbb{R}^{(J\cdot N_{\text{MC}})\times P}$ ($P=64^2$ pixels)
satisfies $G(x)\approx D^{\top}D$, and its truncated singular value
decomposition $D=U\Sigma V^{\top}$ yields $\lambda_k=\Sigma_{kk}^2$
and $v_k=V_{:,k}$ as the top-$50$ eigenvalues and eigenvectors of the
estimated metric.

\paragraph{Sampling responses from the diffused likelihood.}
Evaluating Eq.~\eqref{eq:line} requires drawing
$r\sim p_t(r\mid x_t)$, the response distribution conditional on the
\emph{diffused} image $x_t$. A direct Monte Carlo procedure would
first draw $x\sim p(x\mid x_t)$ and then $r\sim p(r\mid x)$, but
sampling $x\sim p(x\mid x_t)$ requires a full backward pass of the
diffusion model at each $(x_t,t)$ and is prohibitively expensive. We
follow the approximation introduced by \citet{chung2023diffusion}
and adopted for the information-decomposition setting by
\citet{laquitaine2025decomposing}: replace the expensive $x\sim
p(x\mid x_t)$ step by a point mass at the posterior mean
$\hat x(x_t)=\mathbb{E}[x\mid x_t]$, which can be obtained in a
single forward pass of the score network. Concretely, at each
$(x_t,t)$ we proceed in three steps. First, we read the unconditional
posterior mean off Eq.~\eqref{eq:xhat-eps} with $r=\mathbf 0$,
\begin{equation*}
\hat x(x_t) \;=\; \frac{x_t - \sqrt{1-\bar\alpha_t}\,\epsilon^{u}_\theta(x_t,t)}{\sqrt{\bar\alpha_t}},
\end{equation*}
using $\epsilon^{u}_\theta(x_t,t)\approx\epsilon_\theta(x_t,t,\mathbf 0)$
from the null-token prediction of the same network (enabled by the
CFG training of Appendix~\ref{app:diff-model-arch}). Second, we clip
$\hat x(x_t)$ to $[-1,1]$ since the clean image lives in this range.
Third, we draw $r = f_\phi(\hat x(x_t)) + \sigma\xi$ with
$\xi\sim\mathcal{N}(0,I)$ and $\sigma=0.22$, i.e.\ we sample from the
observation model of Eq.~\eqref{eq:obs-model} evaluated at the
denoised image. For the InceptionV3 and ResNet-50 encoders, $f_\phi$ accepts the $[-1,1]$-rescaled image directly; for the ON-OFF encoder, which operates on luminance images in $[0,1]$ (Appendix~\ref{app:encoder-onoff}), a thin wrapper rescales the clipped $\hat x(x_t)\in[-1,1]$ to $[0,1]$ before applying the front-end. The total cost is one extra forward pass of
$\epsilon_\theta$ per Monte Carlo batch to compute
$\epsilon^{u}_\theta(x_t,t)$, plus one evaluation of $f_\phi$ on the
denoised image.

\paragraph{Validity of the point-mass approximation at coarse scales.}
Replacing the broad diffused likelihood $p_t(r\mid x_t)$ by a point mass at the posterior mean $\hat x(x_t)$ collapses the response variance to the observation-noise floor $\sigma^2 I$; the gap is largest at coarse scales (large $\tau$), where the true posterior $p(x\mid x_t)$ is broad. We accept this trade-off because the integrand we are approximating, $J_t(x_t)$, is itself \emph{small} at large $\tau$: as $\tau\to\infty$ the diffused likelihood $p_t(r\mid x_t)$ converges to the marginal $p(r)$, so the true Fisher contribution at coarse scales tends to zero independently of the sampling approximation, and any point-mass bias is multiplied by a vanishing weight. An exact alternative --- drawing $x\sim p(x\mid x_t)$ via a full backward diffusion pass at every Monte Carlo slot --- would be prohibitive at our compute budget, and the point-mass approximation is established practice for diffusion posterior sampling \mbox{\citep{chung2023diffusion}} and in the closely related information-decomposition setting \mbox{\citep{laquitaine2025decomposing}}. The cross-encoder robustness reported in Fig.~\mbox{\ref{fig:4}} (Inception and ResNet-50 encoders fitted independently to the same data) provides indirect evidence that the bias does not dominate the leading $G(x)$ subspace, since the two encoders traverse different bias profiles and yet agree on the leading directions.

\paragraph{Conditional-pair estimator.}
For every $(x_t,t)$ and every Monte Carlo slot we draw \emph{two}
independent response vectors
$r_1,r_2\sim p(r\mid \hat x(x_t))$ and take the difference of their
conditional noise predictions,
\begin{equation}
D_{(j,s), :} \;=\; \sqrt{\tfrac{w_j}{2\,N_{\text{MC}}}}\;\bigl(\epsilon_\theta(x_t,t,r_1)-\epsilon_\theta(x_t,t,r_2)\bigr),
\label{eq:cc}
\end{equation}
where the factor $\tfrac12$ inside the square root compensates for
$\operatorname{Var}(A-B)=2\operatorname{Var}(A)$ under i.i.d.\ draws
and $w_j$ is the log-$\tau$ trapezoid weight introduced above.
Because both terms in Eq.~\eqref{eq:cc} are evaluated at the same
$(x_t,t)$ and use the same conditional cross-section of
$\epsilon_\theta$, any systematic error of the denoiser that is
common to the two draws cancels in the subtraction.
Eq.~\eqref{eq:cc}, the conditional--conditional (CC) form, is an alternative Monte Carlo realisation of
Eq.~\eqref{eq:line} to the unconditional--conditional (UC) form
$\epsilon_\theta(x_t,t,r)-\epsilon^{u}_\theta(x_t,t)$ used by
\citet{laquitaine2025decomposing} with two separately trained
networks; we use Eq.~\eqref{eq:cc} throughout for consistency with
our single-network CFG recipe of
Appendix~\ref{app:diff-model-arch}. On a small held-out set of test images with multiple independent Monte Carlo realisations per image, and at matched cost ($300$ forward passes of the network per scale for each estimator), the top-$5$ eigenvalues of the CC estimate had a median $67\%$ lower Monte Carlo variance than those of the UC form. The leading UC eigenvalue was $2.8$--$5.5$ times the CC value, and this factor varied across images, so it cannot be corrected by a global rescaling of the metric. The main advantage of the CC form is that it avoids this mismatch between conditional and unconditional predictions. Per Monte Carlo sample, CC costs three forward passes (one null-token and two conditional), and UC two.

\paragraph{Algorithm.}
Algorithm~\ref{alg:cc} collects the steps of
Appendix~\ref{app:diff-model-est} into a single pseudocode listing for
reproducibility. A forward pass through $\epsilon_\theta$ consists of
passing the noisy image $x_t$ and diffusion step $t$ through the
UNet2DConditionModel while feeding the response vector $r$ (or the
null token $\mathbf 0$) through the 3-layer SiLU MLP of
Appendix~\ref{app:diff-model-arch}; the resulting $256$-dimensional
token is supplied as the single-token
\texttt{encoder\_hidden\_states} argument to the cross-attention layers of the UNet.

\begin{algorithm}[h]
\caption{Conditional-pair estimator of the top-$k$ eigenpairs of $G(x)$.}
\label{alg:cc}
\begin{algorithmic}[1]
\Require clean image $x\in[-1,1]^{64\times 64}$;
trained score network $\epsilon_\theta$;
encoder $f_\phi$;
DDPM schedule $\{\bar\alpha_t\}_{t=1}^{T}$;
timestep grid $\{t_j\}_{j=1}^{J}$ with $J=24$, $t_1=40$, $t_J=960$;
Monte Carlo count $N_{\text{MC}}=200$;
batch size $B=25$;
observation-noise scale $\sigma=0.22$;
rank $k$
\State Compute $\log\tau_j \gets \log\bigl[(1-\bar\alpha_{t_j})/\bar\alpha_{t_j}\bigr]$ for $j=1,\ldots,J$
\State Compute trapezoid weights $w_j \gets \tfrac12(\log\tau_{j+1}-\log\tau_{j-1})$ (with half-weights at endpoints)
\State Initialise empty list $\mathcal D$
\For{$j \gets 1$ to $J$}
  \State $s_j \gets \sqrt{w_j / (2 N_{\text{MC}})}$
  \For{$n \gets 1$ to $N_{\text{MC}}$ in batches of $B$}
    \State Draw $z \sim \mathcal{N}(0,I)$, shape $(B,1,64,64)$
    \State $x_t \gets \sqrt{\bar\alpha_{t_j}}\, x + \sqrt{1-\bar\alpha_{t_j}}\, z$
      \Comment{Eq.~\eqref{eq:forward-diff}}
    \State $\epsilon^{u} \gets \epsilon_\theta(x_t,\, t_j,\, \mathbf 0)$
      \Comment{null-token forward pass}
    \State $\hat x \gets (x_t - \sqrt{1-\bar\alpha_{t_j}}\,\epsilon^{u}) / \sqrt{\bar\alpha_{t_j}}$
      \Comment{Eq.~\eqref{eq:xhat-eps} with $r=\mathbf 0$}
    \State $\hat x \gets \mathrm{clip}(\hat x,\,-1,\,1)$
    \State $\mu \gets f_\phi(\hat x)$
      \Comment{deterministic response prediction}
    \State Draw $\xi_1, \xi_2 \sim \mathcal{N}(0,I)$, shape $(B, 50)$
    \State $r_1 \gets \mu + \sigma\,\xi_1$;\quad $r_2 \gets \mu + \sigma\,\xi_2$
      \Comment{Eq.~\eqref{eq:obs-model}}
    \State $\epsilon^{c}_1 \gets \epsilon_\theta(x_t,\, t_j,\, r_1)$;\quad $\epsilon^{c}_2 \gets \epsilon_\theta(x_t,\, t_j,\, r_2)$
    \State $\Delta \gets \mathrm{flatten}(\epsilon^{c}_1 - \epsilon^{c}_2)$, shape $(B, 64^2)$
    \State Append $s_j\,\Delta$ to $\mathcal D$
      \Comment{Eq.~\eqref{eq:cc}}
  \EndFor
\EndFor
\State Stack $\mathcal D$ into a matrix $D \in \mathbb{R}^{(J\cdot N_{\text{MC}})\times 64^2}$
\State $(U,\Sigma,V^{\!\top}) \gets \mathrm{SVD}(D)$
\State \Return eigenvalues $\{\Sigma_{\ell\ell}^{2}\}_{\ell=1}^{k}$ and eigenvectors $\{V_{:,\ell}\}_{\ell=1}^{k}$ of $G(x)\approx D^{\!\top}D$
\end{algorithmic}
\end{algorithm}

\paragraph{Fisher baseline.}
The classical Fisher information matrix $J(x)$ used as the comparison baseline in Figs.~\ref{fig:3}F, \ref{fig:3}H, \ref{fig:4}C, \ref{fig:4}E and \ref{fig:5}B is computed in closed form from the Gaussian observation model of Eq.~\eqref{eq:obs-model}. Since $p(r\mid x)=\mathcal{N}(f_\phi(x),\sigma^2 I)$ has constant covariance, the standard formula for the Fisher information of a multivariate Gaussian reduces to
\begin{equation}
J(x) \;=\; \frac{1}{\sigma^2}\, J_{f_\phi}(x)^{\top} J_{f_\phi}(x),
\label{eq:fisher-baseline}
\end{equation}
where $J_{f_\phi}(x)\in\mathbb{R}^{n\times P}$ is the Jacobian of the encoder mean $f_\phi:\mathbb{R}^P\!\to\!\mathbb{R}^{n}$ at the input image $x$, with $P=64\times 64=4096$ pixels and $n=50$ target neurons. We evaluate $J_{f_\phi}(x)$ via PyTorch reverse-mode autograd, materialising it row-by-row (one backward pass per neuron). Because $J_{f_\phi}$ has rank at most $n\ll P$, we then extract the top-$50$ eigenpairs of $J(x)$ directly from the SVD of $J_{f_\phi}=U S V^{\top}$ rather than forming the $P\times P$ outer product: the eigenvalues of $J(x)$ are $S_i^2/\sigma^2$ and the eigenvectors are the rows of $V^{\top}$. The same set of $200$ test images is used for the multi-scale metric, so any Fisher-vs-$G(x)$ comparison fixes the encoder, the dataset, and the rank. The test images are the first $200$ images of the TinyImageNet validation set, which the diffusion models never saw during training.

\section{Predictive neural models \label{app:encoding_mod}}

The goal of the encoding models described in this appendix is to
provide the image-to-response map $f_\phi$ of
Eq.~\eqref{eq:obs-model}: given an image $x$, $f_\phi(x)$ should
predict the trial-averaged \texttt{normMUA} of each recorded neuron
so that the Gaussian observation model
$p(r\mid x) = \mathcal{N}(f_\phi(x),\sigma^2 I)$ is a good fit to the
data. We fit $f_\phi$ by maximising the corresponding Gaussian
log-likelihood on paired stimulus--response data, which under our
model reduces to minimising the per-neuron mean-squared error between
predicted and trial-averaged responses. The fitted encoder then
serves both as a readout of the underlying neural code
(Figs.~\ref{fig:3}--\ref{fig:4}) and as the deterministic component
$f_\phi$ used inside the response sampler of
Appendix~\ref{app:diff-model-est}. This appendix also describes the
Bures--Wasserstein distance on PSD matrices that we use as an
alternative measure of geometric similarity between reconstructed
metric tensors.

\subsection{Data processing}
\label{app:data}

All encoding models are fitted on the \emph{THINGS ventral-stream spiking dataset} (TVSD) from \citet{papale2025extensive}, which reports multi-unit activity
from chronically implanted Utah-array recordings in an awake-behaving
macaque while it fixated on 22{,}248 natural photographs drawn from the
THINGS database \citep{hebart2019things}. The dataset provides a normalised multi-unit firing
rate (\texttt{normMUA}) per recording channel, the train/test split used
in~\citet{papale2025extensive}, and a per-channel split-half \emph{oracle}
reliability that quantifies how much of the trial-averaged response is
reproducible across repeats. Following the original paper, we rely on
the provided electrode-to-area assignment (V1: channels $0$--$511$; V4:
$512$--$767$) and, within each area, retain the $50$ channels with the
highest oracle reliability as the targets of our encoders. For Inception \citep{szegedy2016rethinking} and ResNet \citep{he2016deep} encoders the stimuli are resized to
$64\times 64$, converted to a single grayscale channel, replicated across
the three input channels expected by the pre-trained backbone, and
rescaled to $[-1,1]$ via the usual mean-$0.5$, std-$0.5$ normalisation.
For the ON-OFF encoder, which operates directly on a single-channel
luminance image in $[0,1]$, we omit the final rescaling step.

\begin{table}[h]
\centering
\caption{Predictive performance of all encoders reported in
Figs.~\ref{fig:3}--\ref{fig:4}. For each encoder and area we list the
mean raw Pearson test correlation $\bar r$ between predicted and
trial-averaged neural responses on the Papale test set, and the
corresponding noise-corrected correlation $\bar r_{\text{nc}}$. The
latter is computed per neuron as $r_n/\rho_n$, where $\rho_n$ is the
split-half oracle reliability of neuron $n$, and then averaged across
the $50$ target neurons; values close to $1$ indicate performance near
the intrinsic noise ceiling of the recording.}
\label{tab:encoder_perf}
\footnotesize
\begin{tabular}{lcccc}
\toprule
 & \multicolumn{2}{c}{V1} & \multicolumn{2}{c}{V4} \\
\cmidrule(lr){2-3}\cmidrule(lr){4-5}
Encoder & $\bar r$ & $\bar r_{\text{nc}}$ & $\bar r$ & $\bar r_{\text{nc}}$ \\
\midrule
Inception  & $0.75$ & $0.81$ & $\mathbf{0.71}$ & $\mathbf{0.79}$ \\
ResNet-50  & $\mathbf{0.78}$ & $\mathbf{0.85}$ & $0.70$ & $0.78$ \\
ON-OFF     & $0.20$ & $0.22$ & $0.13$ & $0.14$ \\
\bottomrule
\end{tabular}
\end{table}

Per-neuron oracle reliabilities (mean across the selected 50 neurons)
are $\rho\!=\!0.92$ for V1 and $\rho\!=\!0.90$ for V4. These values put
a ceiling on what \emph{any} encoder can achieve against trial-averaged
responses; the noise-corrected correlations in
Table~\ref{tab:encoder_perf} should therefore be read as the fraction of
that explainable signal actually captured by each encoder.

\subsection{InceptionV3 model}
\label{app:encoder-inception}

Both of our deep CNN encoders follow the factorised-readout template
that is standard in the task-trained vision-to-neural-prediction
literature~\citep{yamins2016using, schrimpf2018brain, papale2025extensive}:
a task-trained classifier backbone is truncated at an intermediate
layer and frozen, and only a small per-neuron readout is fit to the
neural data. For the encoder described in this subsection, the
backbone is InceptionV3~\citep{szegedy2016rethinking} with the
ImageNet-pretrained weights provided by
\texttt{torchvision}~\citep{torchvision2016}, truncated at the
\texttt{Conv2d\_4a\_3x3} layer so that a $64\times 64$ input image
produces a feature map of shape $192\times 12\times 12$.

We chose the truncation layer in a preliminary sweep over the candidate Inception blocks for each area. \texttt{Conv2d\_4a\_3x3} gave the highest correlation in \emph{both} V1 and V4; deeper blocks did not help V4 and hurt V1. This
departs from the common convention of assigning earlier layers to V1 and
deeper layers to V4/IT~\citep{yamins2016using, schrimpf2018brain}, but is
itself common in practice: many brain-score leaderboard entries report a
single best layer across multiple ventral-stream
areas~\citep{schrimpf2018brain}.

Using the same layer for V1 and V4 is important for the
interpretation of our results: because both areas are read out from
an identical feature basis, any systematic difference in $G(x)$
between V1 and V4 (Fig.~\ref{fig:3}) cannot be a layer-choice artefact
and must reflect how the shared features are weighted to predict each
area. The backbone is frozen throughout training.

On top of the frozen features we fit a per-neuron spatial-attention
readout: each target neuron $n$ is parameterised by a centre
$\mu_n\in\mathbb{R}^2$, a width $\sigma_n>0$, a feature-weight vector
$w_n\in\mathbb{R}^C$ and a bias $b_n$. The predicted response is
\begin{equation}
\hat r_n(x) \;=\; \sum_c w_{n,c} \sum_{h,w} a_n(h,w)\, \phi_c(x)_{h,w} \;+\; b_n,
\qquad
a_n(h,w) \propto \exp\!\left(-\tfrac{\|(h,w)-\mu_n\|^2}{2\sigma_n^2}\right),
\label{eq:readout}
\end{equation}
where $a_n$ is normalised to sum to one over the feature-map grid and
$\phi_c(x)$ is the $c$-th channel of the frozen backbone output. We optimise the per-neuron readout with AdamW at learning rate
$3\cdot 10^{-4}$, batch size $64$, dropout $0.25$ applied to the feature
map, selecting the epoch with the highest validation correlation (epochs $79$ and $93$ for V1 and V4). Model selection uses a $10\%$ hold-out
validation split of the THINGS training images, kept disjoint from the
Papale test set; all accuracy numbers used in Figs.~\ref{fig:3}--\ref{fig:4}
and Table~\ref{tab:encoder_perf} are computed on the held-out Papale test
images, averaging the Pearson correlation between predicted and
trial-averaged responses across the $50$ target neurons. Our Inception
encoder reaches $0.81$ (V1) and $0.79$ (V4) of the per-neuron noise
ceiling (Table~\ref{tab:encoder_perf}), matching the headline
performance reported by~\citet{papale2025extensive} for comparable
factorised readouts on TVSD.
\subsection{Resnet50 model}
\label{app:encoder-resnet}

To control for backbone-specific artefacts we also fit an encoder with
the same readout structure on top of a pre-trained ResNet-50 \citep{he2016deep}. Input
images are standardised with the ImageNet mean and standard deviation
inside the backbone wrapper; feature extraction stops at the first block
of the first stage (\texttt{layer1.0}), which produces a $256\times
16\times 16$ feature map for our $64\times 64$ input. The remainder of the readout, the loss, and the optimisation schedule
are identical to those of the Inception encoder. On the Papale test set
the ResNet-50 encoder reaches $0.85$ (V1) and $0.78$ (V4) of the
per-neuron noise ceiling (Table~\ref{tab:encoder_perf}), on
par with or slightly above the Inception encoder on the same neurons.
This matched accuracy is what allows the cross-architecture comparison
in Fig.~\ref{fig:4}: any systematic differences between the two
encoders' geometries that survive cross-model averaging cannot be
attributed to one model being a substantially better fit to the data.

\subsection{On-off model}
\label{app:encoder-onoff}

Our third encoder is a biologically motivated front-end in the spirit
of \citet{berardino2017eigen}. The backbone is the \texttt{OnOff}
module from the \texttt{plenoptic} library
\citep{duong2023plenoptic}, which produces two spatial channels
encoding on-centre and off-centre luminance and contrast responses via
centre-surround filters ($31\times 31$ kernels loaded with the
pre-trained Berardino weights) and is kept frozen throughout training,
exactly as for the Inception and ResNet backbones. The frozen
front-end consists of $12$ biologically interpretable scalars ---
on/off luminance and contrast gain controls and centre/surround radii
of the difference-of-Gaussians filters --- four orders of magnitude
fewer than the $\sim\!173\mathrm{k}$ frozen convolutional parameters
in the Inception backbone (Appendix~\ref{app:encoder-inception}).
On top of the frozen ON-OFF feature map  we fit a dense linear readout that maps the $8{,}192$ features to the $50$ target neurons ($409{,}650$ trainable parameters), by minimising the mean-squared error with model selection on validation correlation.
Test correlations on the Papale test set are substantially lower than
with the deep encoders: the ON-OFF encoder reaches about $22\%$ of the
per-neuron noise ceiling in V1 and $14\%$ in V4
(Table~\ref{tab:encoder_perf}), as expected from the dramatic
reduction in front-end capacity. As discussed in the main text, this
simpler encoder is informative not because it matches the data but
because its Fisher geometry becomes interpretable once the model no
longer has enough capacity to fit high-frequency idiosyncrasies of
the neural code.

\subsection{Comparing different geometries}
\label{app:bures}

Throughout Fig.~\ref{fig:4}D--E we quantify how similar two metrics $A$ and $B$ are by the alignment of their leading eigenvector subspaces. Given truncated eigendecompositions $A\approx V_A\Lambda_A V_A^{\top}$ and $B\approx V_B\Lambda_B V_B^{\top}$ with orthonormal $V_A,V_B\in\mathbb{R}^{P\times k}$, the \emph{subspace alignment} of $A$ and $B$ at rank $k$ is
\begin{equation}
\mathcal{A}_k(A,B) \;=\; \frac{1}{k}\,\|V_A^{\top}V_B\|_F^2 \;=\; \frac{1}{k}\sum_{i=1}^{k}\cos^2\theta_i,
\label{eq:subspace-alignment}
\end{equation}
the mean squared cosine of the principal angles $\{\theta_i\}_{i=1}^{k}$ between the two subspaces \citep{golub2013matrix}; $\mathcal{A}_k\in[0,1]$, with $\mathcal{A}_k=1$ iff the top-$k$ subspaces coincide.

Subspace alignment (Fig.~\ref{fig:4}D--E) only constrains the angles
between eigenvectors and is insensitive to the magnitude of the
associated eigenvalues. To double-check the conclusions of
Fig.~\ref{fig:4} we therefore also compare the reconstructed metric
tensors using the \emph{Bures--Wasserstein} distance \citep{bhatia2019bures} on positive
semi-definite (PSD) matrices, which simultaneously penalises mismatches
in orientation and in scale. Given two PSD matrices $A$ and $B$ the
Bures--Wasserstein distance is
\begin{equation}
d_{\text{BW}}^2(A,B) \;=\; \operatorname{tr}(A)+\operatorname{tr}(B) \;-\; 2\operatorname{tr}\!\Bigl((A^{1/2}BA^{1/2})^{1/2}\Bigr),
\label{eq:bures}
\end{equation}
and corresponds, for symmetric mean-zero Gaussians with covariances $A$
and $B$, to the $2$-Wasserstein distance between the two distributions.
Of the many distances available on the cone of PSD matrices,
\citet{harvey2024representational} argue that Bures--Wasserstein is the
most appropriate choice for \emph{decodable-information} notions of
representational similarity, since it has a direct optimal-transport
interpretation between the two representational geometries.

In practice we work with the truncated eigendecomposition
$A\approx V_A\Lambda_A V_A^{\top}$, $B\approx V_B\Lambda_B V_B^{\top}$
returned by the diffusion estimator of
Appendix~\ref{app:diff-model-est}, with $V_A,V_B\in\mathbb{R}^{P\times k}$ orthonormal and $\Lambda_A$, $\Lambda_B$ the
$\mathrm{diag}$ of the top-$k$ eigenvalues. Writing
$C=V_A^{\top}V_B\in\mathbb{R}^{k\times k}$ for the overlap matrix between the two sets of
eigenvectors, the cross term in Eq.~\eqref{eq:bures} becomes
$\operatorname{tr}\!\Bigl[\bigl(\Lambda_A^{1/2}\,C\,\Lambda_B\,C^{\top}\,\Lambda_A^{1/2}\bigr)^{1/2}\Bigr]=\sum_i \mu_i^{1/2}$
where $\{\mu_i\}$ are the eigenvalues of
$M=\Lambda_A^{1/2}C\Lambda_B C^{\top}\Lambda_A^{1/2}$; we symmetrise $M$
before diagonalisation to guard against numerical asymmetry and sweep
$k=1,\ldots,50$.

\begin{figure}
\begin{center}
\includegraphics[width=1.0\linewidth]{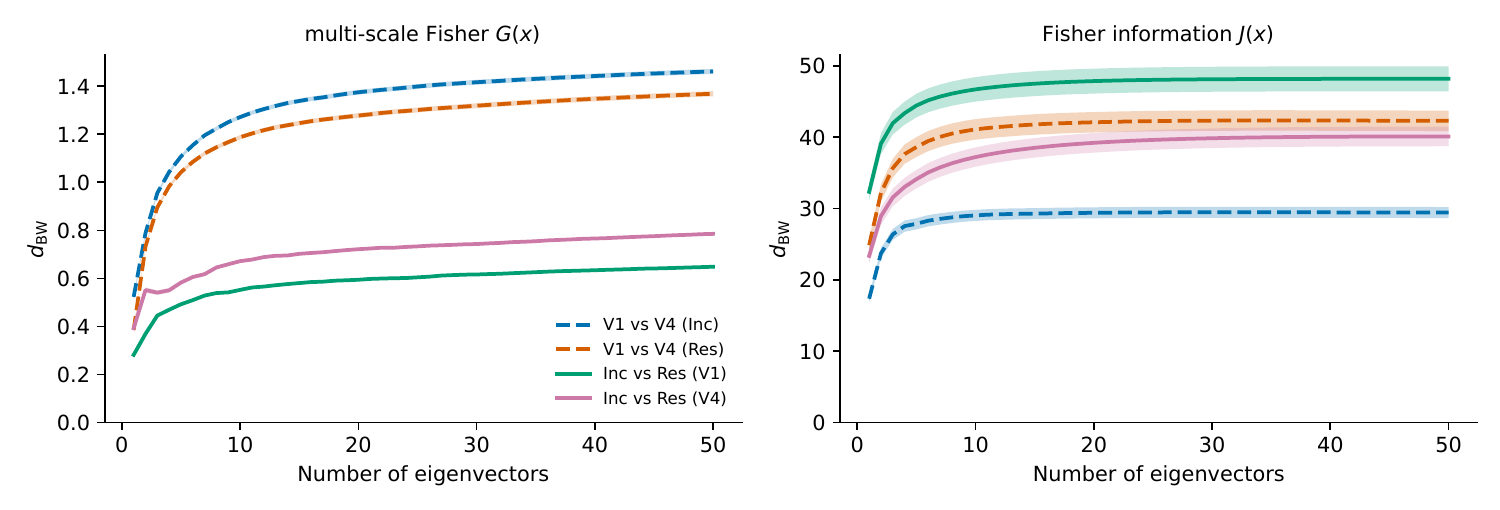}
\caption{\label{fig:app-bw}
\textbf{Bures--Wasserstein distance between reconstructed geometries.}
$d_{\text{BW}}$ computed from the top-$k$ eigenvalue/eigenvector pairs
(Eq.~\eqref{eq:bures}) as a function of $k$, averaged across images,
for four encoder/area comparisons: V1 vs V4 within each backbone
(Inception, ResNet) and Inception vs ResNet within each area (V1, V4).
(\textbf{Left}) Multi-scale metric $G(x)$; cross-model distances (same
area, different backbone) lie below cross-area distances (same backbone,
different area). (\textbf{Right}) Fisher information $J(x)$; the ordering
reverses, with cross-model distances exceeding cross-area distances.
Shaded bands: s.e.m.\ across $200$ test images.
}
\end{center}
\end{figure}

Fig.~\ref{fig:app-bw} reports $d_{\text{BW}}$ as a function of $k$,
averaged across $200$ test images, for the four encoder/area
comparisons of Fig.~\ref{fig:4}. For the multi-scale metric, cross-model distances
(same area, different backbone) lie below cross-area distances (same
backbone, different area); for the Fisher information the ordering
reverses on average (three of the four cross-model/cross-area pairings). This matches the subspace-alignment ordering of
Fig.~\ref{fig:4}D--E, so our conclusions about the multi-scale metric
do not depend on the specific choice of similarity measure.

\end{document}